\documentclass[journal]{IEEEtran}
\usepackage{fancyhdr}
\usepackage{amsmath,amssymb}
\usepackage{amsfonts}
\usepackage[dvips]{graphicx}
\usepackage{dsfont}
\usepackage{comment}

\usepackage[hidelinks]{hyperref}    

\usepackage{color}
\usepackage{pgfplots}           
\pgfplotsset{compat=1.16}
\usepackage{enumerate}
\usepackage{graphics}
\usepackage{subfigure}
\usepackage{cite}
\usepackage{mathrsfs}
\usepackage{stfloats}

\usepgfplotslibrary{fillbetween}
\usetikzlibrary{patterns}
\usetikzlibrary {patterns,patterns.meta}

\usepackage{tikz}
\usepackage{mathdots}
\usepackage{yhmath}
\usepackage{cancel}
\usepackage{color}
\usepackage{siunitx}
\usepackage{array}
\usepackage{multirow}
\usepackage{textcomp}
\usepackage{gensymb}
\usepackage{tabularx}
\usepackage{booktabs}
\usepackage{graphicx}
\usepackage{xcolor}
\usepackage[utf8]{inputenc}
\usepackage[T1]{fontenc}

\newtheorem{theorem}{Theorem}

\renewcommand{\dbinom}[2]{\left(\!\!\begin{array}{c}{#1} \\ {#2} \end{array}\!\!\right)}
\let\oldbinom\binom
\renewcommand{\binom}[2]{\mathchoice{\dbinom{#1}{#2}}{\oldbinom{#1}{#2}}{\oldbinom{#1}{#2}}{\oldbinom{#1}{#2}}}

\allowdisplaybreaks

\begin{document}
\title{Distributed Uplink Rate Splitting Multiple Access (DU-RSMA): Principles and Performance Analysis}
\author{
    Apostolos A. Tegos, Yue Xiao, Sotiris A. Tegos,~\IEEEmembership{Senior Member,~IEEE,}
    \\
    George K. Karagiannidis,~\IEEEmembership{Fellow,~IEEE,} and Panagiotis D. Diamantoulakis,~\IEEEmembership{Senior Member,~IEEE}
    \thanks{A. A. Tegos is with the Department of Electrical and Computer Engineering, Aristotle University of Thessaloniki, 54124 Thessaloniki, Greece (e-mail: apotegath@auth.gr).}
    \thanks{Y. Xiao is with the Provincial Key Laboratory of Information Coding and Transmission, Southwest Jiaotong University, Chengdu 610031, China (e-mail: xiaoyue@swjtu.edu.com)}
    \thanks{S. A. Tegos, G. K. Karagiannidis and P. D. Diamantoulakis are with the Department of Electrical and Computer Engineering, Aristotle University of Thessaloniki, 54124 Thessaloniki, Greece and with the Provincial Key Laboratory of Information Coding and Transmission, Southwest Jiaotong University, Chengdu 610031, China (e-mails: tegosoti@auth.gr, geokarag@auth.gr, padiaman@auth.gr).}
    \thanks{This work was supported by NSFC 62350710217 and Sichuan HT 2024JDHJ0042. The work of Y. Xiao was supported by CPSF 2023TQ0278 amd Postdoctoral Fellowship Program of CPSF-B GZB20230613.}
    \vspace{-7mm}
  }

\maketitle
\begin{abstract}
    One of the main goals of the upcoming sixth-generation (6G) wireless networks is the ability to support higher network density, while ensuring a high quality of service for each user. In this paper, we introduce distributed uplink rate-splitting multiple access (DU-RSMA), define its basic principles, and provide insights into its advantages. Specifically, a system with two remote radio heads (RRHs) and two users is investigated. To improve the performance of the system, we consider that the RRHs can communicate through a feedback link, and thus they are able to decode the received messages either independently or with the assistance of the other RRH, since the decoded information can be shared through the feedback link. It should be noted that this scheme increases the achievable capacity region compared to the known multiple access schemes, which is also evaluated by a novel metric termed ``fill factor''. Both the case of adaptive transmission rates and the case of fixed transmission rates are investigated. To this end, the ergodic rate is investigated to cover the former case, while the outage probability is studied for the latter. Closed-form expressions are derived for both metrics. Finally, the analytical expressions are validated by simulation results, which explicitly show the impact of each parameter on the performance of the system, and prove that the proposed scheme outperforms the corresponding benchmarks.
\end{abstract}

\begin{IEEEkeywords}
    Rate-splitting multiple access (RSMA), ergodic rate, successive interference cancellation, outage probability
\end{IEEEkeywords}

\section{Introduction}
    The main requirements that the upcoming sixth-generation (6G) wireless communication systems are expected to meet are increased data rates for both downlink and uplink, improved reliability, and higher network density \cite{proceedings},\cite{6G}. Meeting these requirements is a necessary condition for extremely reliable low latency communication (eRLLC), further enhanced mobile broadband (FeMBB), and ultra-massive machine-type communication (umMTC). To meet these requirements effectively, it is essential to develop advanced multiple access schemes suitable for the next generation of wireless networks, referred to as next-generation multiple access schemes \cite{NGMA}.
    
    In this direction, non-orthogonal multiple access (NOMA)\cite{NOMA, SANOMA,ChoiNOMA,ApoNOMA} was initially investigated, as it can increase spectral efficiency and connectivity compared to orthogonal multiple access (OMA). However, rate-splitting multiple access (RSMA) \cite{RSMA} has recently become the subject of research in both academia and industry due to its ability to provide a more comprehensive and resilient transmission framework compared to NOMA. In downlink RSMA communication scenarios, the transmitted message is divided in common and private parts at the transmitter in order to deal with interference. The common parts of the message are combined and encoded into a common stream. These streams are transmitted and decoded by multiple users. On the other hand, the private parts are encoded separately and transmitted to their respective users. This part of the split message also causes interference to other users receiving their private message at the same time. This rate splitting technique allows the receiver to decode the common part first and then subtract it from the superimposed signal. By implementing successive interference cancellation (SIC), each user can then decode their private message. Downlink RSMA includes two special cases, i.e. multiuser linear precoding and power-domain NOMA. Taking these into account, RSMA can improve the downlink rate and quality of service. In uplink RSMA communication scenarios, users split their messages into streams and transmit each stream at a predetermined power level. At the receiving end, the remote radio heads (RRHs) use SIC to decode the transmitted messages. 
    For uplink wireless communications, there are specific methods capable of achieving the optimum rate region, namely NOMA with time sharing \cite{TSNOMA2}, joint encoding/decoding \cite{EncodingNOMA}, and RSMA \cite{RSMA_Begin}. 
    NOMA with time sharing requires multiple time slots and synchronization between users, which increase the complexity of the implementation. The joint encoding/decoding scenario of NOMA is also difficult to implement because random codes have high decoding complexity. Considering this, RSMA can significantly reduce the implementation complexity. 

    \subsection{Literature Review}
    It is worth noting that research on RSMA usually considers the downlink scenario, while the uplink scenario has not been investigated as extensively. Furthermore, although single base station (BS) uplink RSMA has been studied in the literature, research on distributed uplink RSMA (DU-RSMA), where distributed remote radio heads (RRHs) serve multiple users within the same resource block, remains largely unexplored. However, with the increasing density of wireless networks and the resulting increase in cell-edge users, novel schemes are needed to extend the achievable rate region of the distributed multiple access channel (DMAC), where receivers in different locations collaborate to improve network performance in interference-limited scenarios. 
    
    In \cite{SumRateMaximization}, the maximization of the sum rate for uplink RSMA is investigated. To achieve this, the users must adjust their transmit power, while the base station (BS) must optimize the decoding order of the messages transmitted by the users. Furthermore, a user pairing algorithm is introduced to reduce the complexity of the above optimization problem. In \cite{YueCognitive}, the authors study the ergodic rate (ER) of cognitive radio (CR) inspired multiple access. Specifically, two protocols based on RSMA and NOMA with SIC are investigated, which attempt to serve a secondary user in a resource block originally dedicated to a primary user without compromising the quality of service (QoS) of the primary user.
    In \cite{Tsiftsiscognitive}, a rate splitting (RS) strategy for uplink CR inspired NOMA systems is proposed. This scheme, which aims to maximize the achievable rate of the secondary user without negatively affecting the outage performance of the primary user, is shown to outperform existing schemes in terms of outage performance. The performance of an imperfection-aware RSMA framework is investigated in \cite{RSMAImperfectSIC}, while considering the coexistence of heterogeneous services in the network. The metrics used to evaluate the performance of the system are the outage probability (OP), the ER, and the system throughput. Simulation results show the impact of imperfect SIC on the system. In \cite{RSMA1order}, the rate-splitting principle was incorporated with uplink NOMA and the OP and the achievable sum rate were studied. However, the authors considered a specific decoding order. Assuming a specific decoding order, \cite{DecodingOrder} showed that both fairness and outage performance are improved by using RS for a two-user uplink scenario. These works were extended in \cite{RSMAallorders}, where the OP of the transmitted messages is derived considering all possible decoding orders at the BS. These expressions were then used to derive the throughput of each source when slotted ALOHA and RSMA are implemented in a random access (RA) network. In addition, \cite{RSMArandomusers} investigates the outage performance of uplink RSMA with randomly deployed users. Two different scheduling schemes are considered, which can either maximize the rate performance or guarantee scheduling fairness, and a new rate fairness-oriented power allocation strategy is proposed. Additionally, with respect to DMAC, the performance of distributed uplink NOMA (DU-NOMA) with adaptive and fixed transmission rates has been investigated in \cite{Pappi} and \cite{DistributedNOMA}, respectively. This approach allows spectrum sharing between neighboring base stations, which cooperate via a feedback link to perform SIC by exchanging decoded messages. An optimal joint user association and decoding order selection scheme was proposed. It was shown that this scheme outperforms DU-NOMA without feedback and OMA schemes, even when the feedback link is not error-free. However, the performance of this scheme in terms of ER was not investigated, which is an important metric for adaptive transmission rate applications. Finally, the pioneering work of \cite{Alouini} identifies DU-NOMA in coordinated multi-point (CoMP) networks as a promising solution for future generations of wireless networks. 
    
    \subsection{Motivation and Contribution}
    Based on the aforementioned technical literature review, and to the best of the authors' knowledge, none of the existing works have investigated a DU-RSMA scenario. However, taking into account that RSMA has been proven to reliably outperform NOMA, that the performance of DU-NOMA with a feedback link has already been investigated and proven to be better than that of DU-NOMA without a feedback link, and the importance of the DMAC in the advancement of wireless networks, it becomes necessary to investigate the performance of DU-RSMA with feedback. This study provides useful theoretical and practical insights into the improvements provided by the feedback link between the two RRHs on the performance of the system. In particular, the contributions of this work can be summarized as follows: 
    \begin{itemize}
        \item We consider a DU-RSMA consisting of two RRHs and two users. We assume that the two RRHs can communicate over an ideal feedback link, i.e., no errors can occur when information is transmitted via the feedback link. Each user transmits its message to both RRHs, while each RRH can decode the transmitted messages either independently or with the help of the other RRH. Under these assumptions, novel closed-form expressions are derived for the OP and the ER of each user and the system. Since a similar scheme has not been investigated, this is the first time that these analytical expressions appear in the literature.
        \item For the case of adaptive transmission rates, we propose a selection criterion according to which the optimal RRH is chosen to decode each message. Based on this criterion, the ER of each user, and thus the ER of the system, is calculated. Specifically, since each user transmits its message to both RRHs, it is assumed for the ER calculation that the optimal RRH, i.e., the one with the higher SINR for the corresponding message at each instance, is selected to decode that message. Following this criterion, it is shown that the achievable ergodic capacity region of the system is significantly larger than the achievable ergodic capacity region if each RRH operated independently.
        \item It is shown for the first time in the existing literature that DU-RSMA expands the achievable rate region compared to DU-NOMA, changing its shape compared to the pentagonal capacity region of the MAC channel and making it more ``squared''. This can be used to increase data rates and fairness between users. We propose a new metric, termed as ``fill factor-$q$'', to evaluate the ``squareness'' of the capacity region, and thus the performance of each scheme in terms of minimum rate, sum rate, and proportional fairness. 
        \item Simulations and numerical results validate the closed-form expressions derived in the theoretical analysis and show in detail the impact of the different system parameters on the performance of the system. Remarkable conclusions about the system and its operation can be extracted, such as the absence of a floor in the OP, regardless of the thresholds for correct message transmission, and the significant gain in the achievable ergodic capacity region. 
    \end{itemize}

\subsection{Structure}
The rest of the paper is organized as follows: In Section II the proposed protocol is described. Section III presents the theoretical analysis for the ergodic rate, while Section IV presents the theoretical analysis for the OP. In Section V, simulation results are presented to verify the derived analytical results and to illustrate the performance of the system, while Section VI concludes the paper.

\section{Proposed Protocol and Achievable Rate Region}
\label{Proposed}
\begin{figure}
    \centering
    \includegraphics[scale = 0.5]{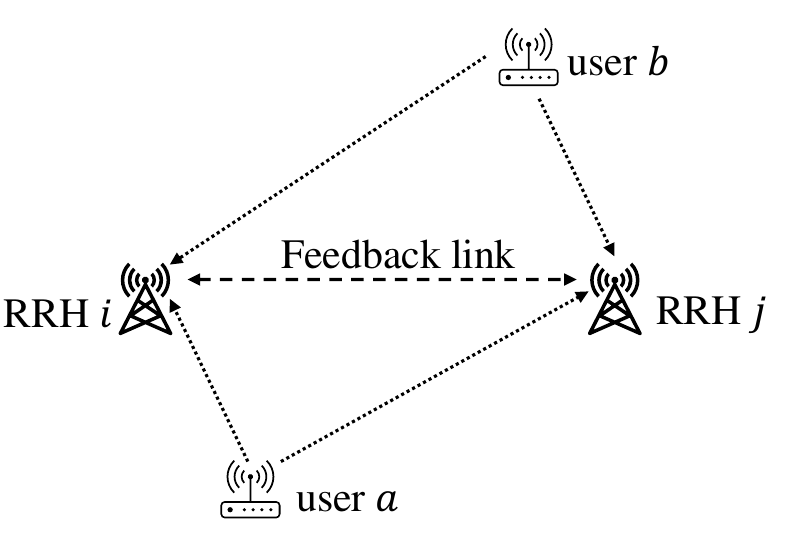}
    \caption{System model.}
    \label{System model}
\end{figure}
We consider a wireless network consisting of two users, $a$ and $b$, and two RRHs, $i$ and $j$. The RRHs can decode the received messages independently, or they can cooperate by exchanging the already decoded messages during SIC, through a feedback link, as shown in Fig. \ref{System model}. In this work, it is assumed that the feedback link is ideal, i.e., the feedback link is error-free. This assumption is realistic and can be achieved using high-capacity fiber optics. It is assumed that all nodes use a single antenna for communication between users and RRHs. 
Without loss of generality, we assume that user $a$ is the one performing message splitting, and thus the message received at RRH $k$ can be expressed as
\begin{equation}
\begin{aligned}
        y_k &= \sqrt{\alpha P_al_{ka}}h_{ka}x_{1a} + \sqrt{\left(1-\alpha\right)P_al_{ka}}h_{ka}x_{2a} \\
        & \qquad + \sqrt{P_bl_{kb}}h_{kb}x_b + n_b,
\end{aligned}
\end{equation}
where $\alpha$ and $1-\alpha$ are the power allocation coefficients for messages $x_{1a}$ and $x_{2a}$, respectively, $h_{km}$ with $k \in \{i,j\}$ and $m \in \{a,b\}$ is the small-scale fading coefficient between the $k$-th RRH and the $m$-th user. Assuming Rayleigh fading, $|h_{km}|^2 \sim \mathcal{CN}(0,1)$. Furthermore, $P_a$, $P_b$ are the powers of each source, $n_b$ is the additive white Gaussian noise (AWGN) with zero mean and variance $\sigma^2$ at the RRHs, and $l_{km}$ denotes the path loss factor between the $k$-th RRH and the $m$-th user, which is given by $l_{km} = cd_{km}^{-n}$, with $c, d_{km},$ and $n$ being the path loss at reference distance $d_0$, the distance between the $k$-th RRH and the $m$-th user, and the path loss exponent, respectively. It has been proven that the optimal decoding order that allows uplink RSMA to reach the capacity region is $\left(x_{1a},x_b,x_{2a}\right)$ \cite{DecodingOrder}. Thus, according to this decoding order, the received signal-to-interference-plus-noise ratio (SINR) at the $k$ RRH for detecting message $x_{1a}$ is given by
\begin{equation}
\label{SNR1a,i}
    \gamma_{1a,k} = \frac{\alpha l_{ka}\gamma_a\lvert h_{ka}\rvert^2}{(1-\alpha)l_{ka}\gamma_a\lvert h_{ka}\rvert^2+l_{kb}\gamma_b\lvert h_{kb}\rvert^2+1},
\end{equation}
where $\gamma_a = \frac{P_a}{\sigma^2}$ and $\gamma_b = \frac{P_b}{\sigma^2}$, are the users' transmit signal-to-noise ratio (SNR), and $k \in \{i,j\}$ as previously stated.
Using SIC, the RRH is able to subtract the decoded signal from the superimposed received signal. Thus, the SINR for messages $x_{b}$ and $x_{2a}$ are given by
\begin{equation}
\label{SNRb,i}
    \gamma_{b,k} = \frac{l_{kb}\gamma_b\lvert h_{kb}\rvert^2}{(1-\alpha)l_{ka}\gamma_a\lvert h_{ka}\rvert^2+1}
\end{equation}
and
\begin{equation}
\label{SNR2a,i}
    \gamma_{2a,k} = (1-\alpha)l_{ka}\gamma_a\lvert h_{ka}\rvert^2.
\end{equation}
Thus, the achievable rates that correspond to each message are given by $R_{x,k}=\log_2(1+\gamma_{x,k})$, with $x\in\{1a,b,2a\}$.

However, in this work, all possible decoding orders are considered to calculate the total OP of each user. When the split messages are transmitted sequentially with appropriate $\alpha$ and $\beta$, i.e., the target rate factor of $x_{1a}$, selection, the performance of RSMA coincides with the performance of NOMA. Therefore, the SINR of NOMA is shown for the remaining decoding orders. In detail, if user $b$ is decoded first, the SINRs for each user for RRH $i$ is given by
\begin{equation}
    \label{SNR-NOMA-ba-a}
    \gamma_{a,k}^{ba} = l_{ka}\gamma_a \lvert h_{ka} \rvert^2, 
\end{equation}
and
\begin{equation}
\label{SNR-NOMA-ba-b}
    \gamma_{b,k}^{ba} = \frac{l_{kb}\gamma_b \lvert h_{kb} \rvert^2}{l_{ka}\gamma_a \lvert h_{ka} \rvert^2+1},
\end{equation}
which are derived by setting $\alpha = 0$ in \eqref{SNR1a,i}-\eqref{SNR2a,i}. On the other hand, if both messages of user $a$ are decoded first the SINRs are given by
\begin{equation}
    \label{SNR-NOMA-ab-a}
    \gamma_{a,k}^{ab} = \frac{l_{ka}\gamma_a \lvert h_{ka} \rvert^2}{l_{kb}\gamma_b \lvert h_{kb} \rvert^2+1},
\end{equation}
and
\begin{equation}
    \label{SNR-NOMA-ab-b}
    \gamma_{b,k}^{ab} = l_{kb}\gamma_b \lvert h_{kb} \rvert^2,
\end{equation}
which is equivalent to setting $\alpha = 1$ in \eqref{SNR1a,i}-\eqref{SNR2a,i}.
\begin{figure}
   \centering
   \begin{tikzpicture}
   \begin{axis}[
   width = 0.85\linewidth,
   xlabel = {Rate user $a$ (bps/Hz)},
   ylabel = {Rate user $b$ (bps/Hz)},
   ymin = 0,
   ymax = 7,
   xmin = 0,
   xmax = 8,
   ytick = {1,3,...,7},
   grid = major,
   legend entries = {{Proposed},{},{},{RRH $i$},{},{RRH $j$},{},{DU-NOMA-TS},{},{},{},{},{},{},{},{},{},{Area A}, {Area B}},
   legend cell align = {left},
   legend style = {font = \scriptsize},
   legend style={at={(0,0)},anchor=south west},
   legend image post style={scale=0.7}
   ]
   \addplot[
    black,
    mark = square,
    mark repeat = 2,
    mark size = 3,
    mark phase = 0,
    no marks,
    line width = 1pt
    ]
    table {Data/ErgodicRate/FixedH/SumErgodic.dat};
    \addplot[
    black,
    mark = square,
    mark repeat = 2,
    mark size = 3,
    mark phase = 0,
    no marks,
    line width = 1pt
    ]
    table {Data/ErgodicRate/FixedH/x1.dat};
    \addplot[
    black,
    mark = square,
    mark repeat = 2,
    mark size = 3,
    mark phase = 0,
    no marks,
    line width = 1pt
    ]
    table {Data/ErgodicRate/FixedH/y1.dat};
    \addplot[
    red,
    mark = square,
    mark repeat = 2,
    mark size = 3,
    mark phase = 0,
    no marks,
    line width = 1pt
    ]
    table {Data/ErgodicRate/FixedH/IRRH.dat};
    \addplot[
    red,
    mark = square,
    mark repeat = 2,
    mark size = 3,
    mark phase = 0,
    no marks,
    line width = 1pt
    ]
    table {Data/ErgodicRate/FixedH/x2.dat};
    \addplot[
    blue,
    mark = square,
    mark repeat = 2,
    mark size = 3,
    mark phase = 0,
    no marks,
    line width = 1pt
    ]
    table {Data/ErgodicRate/FixedH/JRRH.dat};
    \addplot[
    blue,
    mark = square,
    mark repeat = 2,
    mark size = 3,
    mark phase = 0,
    no marks,
    line width = 1pt
    ]
    table {Data/ErgodicRate/FixedH/y2.dat};
    \addplot[
    green,
    mark = square,
    mark repeat = 2,
    mark size = 3,
    mark phase = 0,
    no marks,
    line width = 1pt
    ]
    table {Data/ErgodicRate/FixedH/TS.dat};
    \addplot[
    black,
    mark = square,
    mark repeat = 2,
    mark size = 3,
    mark phase = 0,
    no marks,
    style = dashed,
    line width = 1pt
    ]
    table {Data/ErgodicRate/FixedH/y3.dat};
     \addplot[
    black,
    mark = square,
    mark repeat = 2,
    mark size = 3,
    mark phase = 0,
    no marks,
    style = dashed,
    line width = 1pt
    ]
    table {Data/ErgodicRate/FixedH/x3.dat};
    \addplot[
    black,
    mark = square,
    mark repeat = 2,
    mark size = 3,
    mark phase = 0,
    no marks,
    style = dotted,
    line width = 1pt
    ]
    table {Data/ErgodicRate/FixedH/x4.dat};
    \addplot[
    black,
    mark = square,
    mark repeat = 2,
    mark size = 3,
    mark phase = 0,
    no marks,
    style = dotted,
    line width = 1pt
    ]
    table {Data/ErgodicRate/FixedH/y4.dat};
    \addplot[
    black,
    mark = square,
    mark repeat = 2,
    mark size = 3,
    mark phase = 0,
    no marks,
    style = dashdotted,
    line width = 1pt
    ]
    table {Data/ErgodicRate/FixedH/x5.dat};
    \addplot[
    black,
    mark = square,
    mark repeat = 2,
    mark size = 3,
    mark phase = 0,
    no marks,
    style = dashdotted,
    line width = 1pt
    ]
    table {Data/ErgodicRate/FixedH/y5.dat};
    \addplot[
    green,
    mark = square,
    mark repeat = 2,
    mark size = 3,
    mark phase = 0,
    no marks,
    style = dashdotted,
    line width = 1pt
    ]
    table {Data/ErgodicRate/FixedH/x6.dat};
    \addplot[
    green,
    mark = square,
    mark repeat = 2,
    mark size = 3,
    mark phase = 0,
    no marks,
    style = dashdotted,
    line width = 1pt
    ]
    table {Data/ErgodicRate/FixedH/y6.dat};
    \addplot[only marks, mark=*] coordinates {(6.1749,4.4458) (7.62065, 5.3120443)};
    \node[] at (axis cs: 6.5,4.9){$(R_a^*,R_b^*)$};
    \node[] at (axis cs: 7,5.6){($A,B$)};
    \node[] at (axis cs: 4,6.3){$\mathrm{FF}_2^{\mathrm{DU-RSMA}} = \frac{\text{Area A}}{\text{Area B}}=\frac{R_a^*R_b^*}{AB}$};
\path[name path=lower1] (axis cs:0, 0) -- (axis cs: 6.1749, 0);
\path[name path=upper1] (axis cs:0, 4.4458) -- (axis cs: 6.1749, 4.4458);
\addplot[pattern=north west lines,pattern color=black]
fill between[
of = lower1 and upper1,
soft clip = {domain=0:6.1749}
];
\path[name path=lower2] (axis cs:0, 0) -- (axis cs: 7.62065, 0);
\path[name path=upper2] (axis cs:0, 5.3120443) -- (axis cs: 7.62065, 5.3120443);
\addplot[pattern=dots,pattern color=black]
fill between[
of = lower2 and upper2,
soft clip = {domain=0:7.62065}
];
\addplot[only marks, mark=*, color = green] coordinates {(5.3168,3.5635)};
           \end{axis}
       \end{tikzpicture}
       \caption{Capacity region.}
   \label{Capacity Region}
\end{figure}
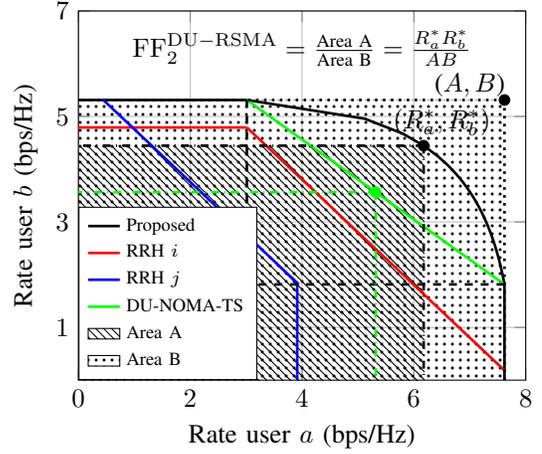
Considering that in the proposed scheme the messages of each user are received by both RRHs, we propose a selection criterion according to which each message is decoded by the best RRH. It should be clarified that both RRHs have perfect channel state information, and before each transmission, the users are informed which $\alpha$ should be used by the user splitting its message. In particular, since the decoding order which allows uplink RSMA to achieve the capacity region is $\left(x_{1a},x_b,x_{2a}\right)$, $\gamma_{1a,i}$ is compared with $\gamma_{1a,j}$ and the RRH with the higher SINR is selected to decode $x_{1a}$. Next, the message from user $b$ is decoded by RRH $i$ if $\gamma_{b,i}>\gamma_{b,j}$, otherwise by RRH $j$. Finally, using a similar comparison between $\gamma_{2a,i}$ and $\gamma_{2a,j}$, the second message of user $a$, i.e., $x_{2a}$, is decoded by the best RRH. 

To derive the capacity region shown in Fig. \ref{Capacity Region}, the above procedure must be repeated for $\alpha \in [0,1]$. On the other hand, the capacity regions of RRH $i$ and RRH $j$ are obtained when each message is decoded by RRH $i$ and RRH $j$, respectively, regardless of the relationship between their corresponding SINRs. As for time-sharing NOMA (TS-NOMA), it is one of the known methods capable of achieving the optimal rate region for wireless uplink communication. Its function is based on allocating part of the time slot to the first decoding configuration (which includes the decoding order and the selection of the RRHs that decode the users' messages) and the rest to the second. However, as shown in Fig. \ref{Capacity Region}, the use of DU-RSMA and the proposed decoding strategy significantly extends the capacity region compared to DU-NOMA with time-sharing studied in \cite{Pappi}. This gain, which is due to the fact that the achievable rate region of DU-RSMA is more ``squared" than that of DU-NOMA, can also be used to increase data rates and fairness between users.  To evaluate the gain of extending the capacity region in terms of minimum rate, sum rate, and proportional fairness, a novel metric is introduced, termed as ``fill factor-$q$''\footnote{The name of this metric was inspired by its use in characterizing the efficiency of solar cells. In solar cells, the fill factor measures ``squareness'' by representing the area of the largest rectangle that fits within the current-voltage curve. Similarly, it quantifies the ``squareness'' of the capacity region.}, which for different values of $q$, is defined as
\begin{equation}
\label{fill_factor_definition}
\begin{split}
&\mathrm{FF}_q=\\
&\frac{f_q(R_a^*,R_b^*)}{f_q(\log_2(1+\max(\gamma_{a,i},\gamma_{a,j})),\log_2(1+\max(\gamma_{b,i},\gamma_{b,j})))}.
\end{split}
\end{equation}
In \eqref{fill_factor_definition}, $R_a^*,R_b^*$ is the pair of rates that maximizes $f_q$, while $f_q$ is given by
\begin{equation}
f_q(x,y)=
\begin{cases}
\min(x,y), &q=0\\
x+y, &q=1\\
xy,&q=2.
\end{cases}
\end{equation}
More specifically, $\mathrm{FF}_0$ is defined as the area of the largest square that fits in the capacity region to the hypothetical largest square that would be achieved if each user were the only one served by the two RRHs (e.g., by using double orthogonal resources) and the rates of the two users were enforced to be symmetrical. Similarly, $\mathrm{FF}_1$ is defined as the perimeter of the rectangle with the largest perimeter that fits in the capacity region to the perimeter that would be achieved if each user were the only one served by the two RRHs. Finally, $\mathrm{FF}_2$ is directly related to the proportional fairness, which is defined as $\log_2(R_a) +\log_2(R_b)=\log_2(R_aR_b)$ \cite{Proportional_Fairness}, and is defined as the area of the largest rectangle that fits in the capacity region to the hypothetical area that would be achieved if each user were the only one served by the two RRHs.


It is obvious that $\mathrm{FF}_q^{\mathrm{DU-RSMA}}>\mathrm{FF}_q^{\mathrm{DU-NOMA}}$. Indicatively, for the case shown in Fig. \ref{Capacity Region}, it can be easily calculated that $\mathrm{FF}_0^{\mathrm{DU-RSMA}} = 0.8789$, $\mathrm{FF}_0^{\mathrm{DU-NOMA}} = 0.6614$, $\mathrm{FF}_1^{\mathrm{DU-RSMA}} = 0.8260$, $\mathrm{FF}_1^{\mathrm{DU-NOMA}} = 0.7290$, $\mathrm{FF}_2^{\mathrm{DU-RSMA}} = 0.6782$, and $\mathrm{FF}_2^{\mathrm{DU-NOMA}} = 0.4680$. This figure is derived for transmit SNR = $75$ dB, $d_{ia} = d_{jb} = 10$ m, $d_{ja} = d_{ja} = 30$ m, $h_{ia} = 1.36$, , $h_{ib} = 0.725$, , $h_{ja} = 2.082$, and $h_{jb} = 1.013$, and assuming that the path loss factor is given by $l_{km} = cd_{km}^{-n}$ with $c = 10^{-3}$ and $n = 2.5$. Note that the point that maximizes the FF of each scheme is not the same as the point that maximizes the sum rate of the system.

\section{Ergodic Rate Analysis}

In scenarios where the target data rate changes dynamically depending on the channel conditions of the source, the ER is a widely used and appropriate metric to evaluate the system performance. Using the above expressions for the SINR, closed-form expressions are derived for the ER of each user as well as the sum ER of the system. As already mentioned, only one decoding order, $\left(x_{1a},x_b,x_{2a}\right)$, is considered when investigating the ER. 
Taking this into account, we derive closed-form expressions for each user's ER.

\subsection{ER of User $a$}
It is known that the ER of user $a$ performing rate splitting is given as
\begin{equation}
    \label{Ergodic Split}
    \begin{aligned}
    \Bar{C}_a = \mathbb{E}\{\log_2(1+\gamma_{1a,k})\} + \mathbb{E}\{\log_2(1+\gamma_{2a,k})\},
    \end{aligned}
\end{equation}
where $\mathbb{E}\{\cdot\}$ denotes the statistical expectation, and $\gamma_{1a,k},\gamma_{2a,k}$ are given by \eqref{SNR1a,i}, and \eqref{SNR2a,i}, respectively, depending on the value of $k$. 
\begin{theorem}
\label{th3}
    In the proposed scheme, the ER of user $a$ is given by
    \begin{equation}
    \begin{aligned}
        \Bar{C}_a = P_1\Bar{C}_{1a}^i + (1-P_1)\Bar{C}_{1a}^j + P_2\Bar{C}_{2a}^i + P_2\Bar{C}_{2a}^j, 
    \end{aligned}
    \end{equation}
    where the expressions for $\Bar{C}_{1a}^i$ is provided at the beginning of the next page and
    \begin{equation}
    \label{Ergodic 2A}
    \Bar{C}_{2a}^i = \frac{-e^\frac{1}{(1-\alpha) l_{ia}\gamma_a}\mathrm{Ei}\left(\frac{1}{(1-\alpha) l_{ia}\gamma_a}\right)}{\ln2}.
    \end{equation}
    The expressions for $\Bar{C}_{1a}^j$ and $\Bar{C}_{2a}^j$ are derived by changing $i$ to $j$ in \eqref{Ergodic 1A} and \eqref{Ergodic 2A}, respectively, where $\mathrm{Ei}(\cdot)$ denotes the exponential integral function \cite[(8.211.1)]{gradshteyn2014table}. $P_1$ denotes the probability that $\gamma_{1a,i}^H>\gamma_{1a,j}^H$ and $P_2$ denotes the probability that $\gamma_{2a,i}^H > \gamma_{2a,j}^H$, and are given by 
    \begin{equation}
    \label{P_1}
    \begin{aligned}
        P_1 = \frac{ l_{ia} l_{jb}}{ l_{ia} l_{jb}- l_{ja} l_{ib}} \left(1+\frac{ l_{ja} l_{ib}\log_2{\frac{ l_{ja} l_{ib}}{ l_{ia} l_{jb}}}}{ l_{ia} l_{jb}- l_{ja} l_{ib}}\right),
        \end{aligned}
    \end{equation}
    and
    \begin{equation}
        P_2 = \frac{ l_{ia}}{ l_{ia}+ l_{ja}}.
    \end{equation}
\end{theorem}
\begin{figure*}
\begin{equation}
\label{Ergodic 1A}
    \Bar{C}_{1a}^i \!=\! \begin{cases}
        \frac{1}{\ln2}\left(\frac{e^{\frac{1}{ l_{ia}\gamma_a}}\mathrm{Ei}\left(-\frac{1}{ l_{ia}\gamma_a}\right)}{ l_{ia}\gamma_a}+1-\frac{(1-\alpha)e^{\frac{1}{(1-\alpha) l_{ia}\gamma_a}}\mathrm{Ei}\left(-\frac{1}{(1-\alpha) l_{ia}\gamma_a}\right)}{\alpha}+\frac{(1-\alpha)e^{\frac{1}{ l_{ia}\gamma_a}}\mathrm{Ei}\left(-\frac{1}{ l_{ia}\gamma_a}\right)}{\alpha}\right), &\frac{ l_{ib}\gamma_b}{ l_{ia}\gamma_a} = 1 \\
        \frac{1}{\ln2}\left(\frac{e^{\frac{1}{(1-\alpha) l_{ia}\gamma_a}}\mathrm{Ei}\left(-\frac{1}{(1-\alpha) l_{ia}\gamma_a}\right)}{\alpha}-\frac{e^{\frac{1}{ l_{ia}\gamma_a}}\mathrm{Ei}\left(-\frac{1}{ l_{ia}\gamma_a}\right)}{\alpha}-\frac{1}{1-\alpha}\left(\frac{e^{\frac{1}{(1-\alpha) l_{ia}\gamma_a}}\mathrm{Ei}\left(-\frac{1}{(1-\alpha) l_{ia}\gamma_a}\right)}{ l_{ia}\gamma_a}+1-\alpha\right)\right), &\frac{ l_{ib}\gamma_b}{ l_{ia}\gamma_a} = 1-\alpha\\
        -\frac{\alpha l_{ia}\gamma_a}{\ln2}\left(\frac{e^{\frac{1}{ l_{ib}\gamma_b}}\mathrm{Ei}\left(-\frac{1}{ l_{ib}\gamma_b}\right)}{ l_{ib}\left(1-\frac{ l_{ia}}{ l_{ib}}\right)\left(1-\frac{(1-a) l_{ia}}{ l_{ib}}\right)}+\frac{e^{\frac{1}{ l_{ia}\gamma_a}}\mathrm{Ei}\left(-\frac{1}{ l_{ia}\gamma_a}\right)}{\alpha( l_{ia}- l_{ib})}+\frac{e^{\frac{1}{(1-\alpha) l_{ia}\gamma_a}}\mathrm{Ei}\left(-\frac{1}{(1-\alpha) l_{ia}\gamma_a}\right)}{(1-\alpha)\left(1-\frac{1}{1-\alpha}\right)\left( l_{ia}-\frac{ l_{ib}}{1-\alpha}\right)}\right), &\text{otherwise}
    \end{cases} 
\end{equation}
\hrulefill
\end{figure*}
\begin{IEEEproof}
The proof is provided in Appendix \ref{Proof3} 
\end{IEEEproof}
\subsection{ER of User $b$}
Regarding user $b$ the ER analysis is more straightforward. The following theorem provides the ER of user $b$.
\begin{theorem}
    In the proposed scheme, the ER of user $b$ is given by
    \begin{equation}
    \label{ER user b}
        \Bar{C}_{b} = (1-P_1)\Bar{C}_{b}^i + P_1\Bar{C}_{b}^j,
    \end{equation}
    where $P_1$ is given by \eqref{P_1} and
    \begin{equation}
\label{ErgodicB}
    \Bar{C}_{b}^i = \begin{cases}
        -\frac{ l_{ib}\gamma_b}{\ln2}\left(\frac{e^{\frac{1}{ l_{ib}\gamma_b}}\mathrm{Ei}\left(-\frac{1}{ l_{ib}\gamma_b}\right)}{ l_{ib}\gamma_b-(1-\alpha) l_{ia}\gamma_a}\right. \\ \left. \quad +\frac{e^{\frac{1}{(1-\alpha) l_{ia}\gamma_a}}\mathrm{Ei}\left(-\frac{1}{(1-\alpha) l_{ia}\gamma_a}\right)}{(1-\alpha) l_{ia}\gamma_a- l_{ib}}\right), &\frac{(1-\alpha) l_{ia}\gamma_a}{ l_{ib}\gamma_b} \neq 1\\
        \frac{ l_{ib}\gamma_b}{\ln2}\left(\frac{e^{\frac{1}{ l_{ib}\gamma_b}}\mathrm{Ei}\left(-\frac{1}{ l_{ib}\gamma_b}\right)}{ l_{ib}\gamma_b}+1\right), &\frac{(1-\alpha) l_{ia}\gamma_a}{ l_{ib}\gamma_b} = 1.
    \end{cases}
\end{equation}
The expression for $\Bar{C}_{b}^j$ is derived by interchanging $i$ with $j$.
\end{theorem}

\begin{IEEEproof}
    Given that user $b$ transmits its message to both RRHs $i$ and $j$, its ER is given by 
    \begin{equation}
    \label{Ergodic User B}
    \begin{aligned}
    \Bar{C}_b = \mathbb{E}\{\max\left(\log_2(1+\gamma_{b,i}),\log_2(1+\gamma_{b,j})\right)\}.
    \end{aligned}
\end{equation}
    According to the above mentioned criterion, the message from user $b$ is decoded by RRH $i$, when $\gamma_{b,i}^H > \gamma_{b,j}^H$, and by RRH $j$ otherwise. Similarly to the previous analysis we get
    \begin{equation}
    \begin{aligned}
        \Pr\left(\gamma_{b,i}^H > \gamma_{b,j}^H\right)
        =\Pr\left(\lvert h_{ia}\rvert^2\lvert h_{jb}\rvert^2 < \frac{l_{ja}l_{ib}}{l_{ia}l_{jb}}\lvert h_{ja}\rvert^2\lvert h_{ib}\rvert^2\right),
    \end{aligned}
    \end{equation}
    which is the complementary event of \eqref{Channel Comparison 1}, and thus the probability that RRH $i$ is chosen for message $x_b$ is $1-P_1$. To calculate, the ER for each RRH we will use
    \begin{equation}
\label{Capacity b,i}
    \Bar{C}_{b}^i = \frac{1}{\ln2}\int_0^\infty\frac{1-F_{\gamma_{b,i}}(x)}{1+x}\mathrm{d}x.
\end{equation}
    From \eqref{Capacity b,i}, it is apparent that the calculation of $F_{\gamma_{b,i}}(x)$ is necessary to proceed. This can be achieved as
    \begin{equation}
    \label{CDF XB}
    \begin{aligned}
        F_{\gamma_{b,i}}(x) &= \Pr\left(\gamma_{b,i} < x\right) \\
        &= \Pr\left(\lvert h_{ib} \rvert^2 < \frac{(1-\alpha)l_{ia}\gamma_a\lvert h_{ia} \rvert^2x+x}{l_{ib}\gamma_b}\right)\\
        &= 1 - \frac{ l_{ib}\gamma_be^{-\frac{x}{ l_{ib}\gamma_b}}}{(1-\alpha) l_{ia}\gamma_a+ l_{ib}\gamma_b}.
        \end{aligned}
    \end{equation}
    Substituting \eqref{CDF XB} in \eqref{Capacity b,i}, it occurs that the ER for RRH $i$ is given by
    \begin{equation}
    \label{Capacity b,i2}
        \Bar{C}_{b}^i = \frac{ l_{ib}\gamma_b}{\ln2}\int_0^\infty\frac{e^{-\frac{x}{ l_{ib}\gamma_b}}}{( l_{ib}\gamma_b+(1-\alpha) l_{ia}\gamma_ax)(1+x)}.
    \end{equation}
    By applying partial fraction decomposition in \eqref{Capacity b,i2} and using \cite[(3.352.4)]{gradshteyn2014table}, the final expression for $\Bar{C}_{b}^i$ is derived, while $\Bar{C}_{b}^j$ can be calculated similarly, which concludes the proof.
\end{IEEEproof}

The sum ER of the system is given by 
\begin{equation}
    \Bar{C} = \Bar{C}_a + \Bar{C}_b.
\end{equation}

\renewcommand{\arraystretch}{1.5}
\begin{table}
    \caption{Used functions}
    \vspace{-3mm}
    \label{HelpingFunctions}
    \centering
    \setlength{\tabcolsep}{8pt}
    \begin{tabular}{c|c}
    \hline \hline
         Term
         &  Expression
         \\ \hline \hline
         $f_1(x)$
         &$\frac{\theta_2 \alpha}{x \gamma_b (\alpha-\theta_{11}(1-\alpha)(1+\theta_2))}$
         \\ \hline
         $f_2(x)$
         &$\frac{\alpha\theta_{12}-\theta_{11}(1-\alpha)(\theta_{12}+1)}{x \gamma_b\theta_{11}(1-\alpha)}$
         \\ \hline
         $f_3(x)$
         &$\frac{\theta_2(\theta_{12}+1)}{x \gamma_b}$
         \\ \hline
         $f_4(x)$
         &$\frac{\theta_2(1+\theta_{1}(1-\alpha))}{x \gamma_b(1-\theta_1\theta_2(1-\alpha))}$
         \\ \hline
         $f_5(x)$
         &$\frac{\theta_{12}-\theta_1(1-\alpha)}{x \gamma_b\theta_1(1-\alpha)}$
         \\ \hline
         $f_6(x)$
         &$\frac{\theta_2\alpha(1+\theta_{11})}{x \gamma_b(\alpha-\theta_{11}(1-\alpha+\theta_2))}$
         \\ \hline
         $f_7(x)$
         &$\frac{\theta_{12}(\alpha-\theta_{11}(1-\alpha))-\theta_{11}(1-\alpha)}{x \gamma_b \theta_{11}(1-\alpha)}$
         \\ \hline
         $f_8(x)$
         &$\frac{\theta_2(\theta_{12}+1-\alpha)}{x \gamma_b(1-\alpha)}$
         \\ \hline
         $f_9(x)$
         &$\frac{\theta_2(\theta_{1}+1)}{x \gamma_b(1-\theta_1\theta_2)}$
         \\ \hline
         $F_1$
         &$-(\alpha-\theta_{11}(1-\alpha)(1+\theta_2))$
         \\ \hline
         $F_2$
         &$-(\alpha-\theta_{11}(1-\alpha+\theta_2))$
         \\ \hline
         $F_3$
         &$1-\theta_1\theta_2(1-\alpha)$ \\
         \hline
         $\Phi_1(x,y,z)$
         &$\frac{x\gamma_a(a-\theta_{11}(1-a))e^{-\frac{\theta_{11}y\gamma_bz+\theta_{11}}{x\gamma_a(a-\theta_{11}(1-a))}-z}}{x\gamma_a(a-\theta_{11}(1-a))+\gamma_by\theta_{11}}$ \\
         \hline
         $\Phi_2(x,y,z)$
         &$\frac{x\theta_2\gamma_a(1-a)e^{-\frac{y\gamma_bz-\theta_2}{x\gamma_a\theta_2(1-a)}-z}}{x\theta_2\gamma_a(1-a)+\gamma_by}$ \\
         \hline
         $\Phi_3(x,y,z)$
         &$\frac{x\theta_2\gamma_ae^{-\frac{y\gamma_bz-\theta_2}{x\gamma_a\theta_2}-z}}{x\theta_2\gamma_a+\gamma_by}$ \\
         \hline
         $\Phi_4(x,y,z)$
         &$\frac{x\gamma_ae^{-\frac{\theta_{1}y\gamma_bz+\theta_{1}}{x\gamma_a}-z}}{xy\gamma_a+\gamma_by\theta_1}$ \\
          \hline \hline
    \end{tabular}
    \vspace{-3mm}
\end{table}

\section{Outage Analysis}
In this section, the derivation of closed-form expressions for the OP of all messages is presented, as OP is an important metric for evaluating system performance, especially when the data rate is fixed. As already mentioned, the derivation of the OP takes into account all possible decoding orders. For the proper functioning of the scheme, a 1-bit feedback, denoted by $z$, is provided to user $a$ (the one performing message splitting) based on the instantaneous channel gains known by the RRHs. Specifically, without loss of generality, we assume that when $z=0$ is transmitted, the information conveyed is that the transmitted messages can be decoded using NOMA, whereas when $z=1$ is transmitted, the information conveyed is that RSMA is the only way to decode the transmitted message. Depending on the feedback,  $\alpha$ is selected, such as $\alpha=0$ if $z=0$ and $\alpha=\alpha_{p}$ if $z=1$, where $\alpha_{p}$ is a predetermined value of $\alpha$ and does not depend on the instantaneous channel gains. It should be clarified that $\alpha = 0$ implies that user $a$ transmits a single message instead of splitting its message into two parts. This does not affect the decoding order of NOMA, which is decided by the BS and is independent of the value of $\alpha$. To derive the OP, we consider predefined thresholds, given as $\theta_{11} = 2^{\beta R_a}-1$, $\theta_{12} = 2^{(1-\beta) R_a}-1$, and $\theta_2 = 2^{R_b}-1$ for messages $\left(x_{1a},x_{2a},x_b\right)$, and $\theta_1 = 2^{R_a}-1$, $\theta_2$ for messages $x_a, x_b$ when NOMA is implemented, where $R_a$ and $R_b$ are the target rates for each user, and $\beta \in [0,1]$ denotes the target rate factor and does not depend on the instantaneous channel gains. It should be noted that these thresholds are common to both RRHs. To keep the closed-form expressions as compact as possible, we define the functions in Table \ref{HelpingFunctions}.

\begin{table*}
    \caption{Expressions for $G_1$}
    \vspace{-3mm}
\label{case G1}
\centering
\setlength{\tabcolsep}{4pt}
\begin{tabular}{c|c|c|c|c}
\hline \hline
\multicolumn{3}{c|}{Conditions}
&Term
& Expression
\\
\hline \hline
\multirow{2}{*}{$\begin{aligned} &\frac{\alpha}{1-\alpha}>\theta_{11} \ \cap \ \\ &\theta_{11} -\theta_1(\alpha-\theta_{11}(1-\alpha))<0\end{aligned}$}
&$F_2 > 0$
&$f_6(l_{ib}) > \frac{\theta_2}{l_{ib}\gamma_b}$
&\multirow{2}{*}[-0.25cm]{$G_{11}^i$}
&$\begin{aligned}
    &1+ \Phi_1\left(l_{ia},l_{ib},f_6(l_{ib})\right)- \Phi_1\left(l_{ia},l_{ib},0\right)\\
    &-e^{-\frac{\theta_2}{ l_{ib}\gamma_b}}- \Phi_3\left(l_{ia},l_{ib},f_6(l_{ib})\right)+ \Phi_3\left(l_{ia},l_{ib},\frac{\theta_2}{l_{ib}\gamma_b}\right)
\end{aligned}$
\\ \cline{2-3} \cline{5-5}
&$F_2 < 0$ 
&
&
&$1- \Phi_1\left(l_{ia},l_{ib},0\right)-e^{-\frac{\theta_2}{ l_{ib}\gamma_b}}+ \Phi_3\left(l_{ia},l_{ib},\frac{\theta_2}{l_{ib}\gamma_b}\right)$
\\ \hline
\multirow{2}{*}{$\begin{aligned} &\frac{\alpha}{1-\alpha}<\theta_{11} \ \cup \ \\&\theta_{11}-\theta_1(\alpha-\theta_{11}(1-\alpha))>0\end{aligned}$}
&$1-\theta_1\theta_2>0$
&$f_9(l_{ib}) > \frac{\theta_2}{l_{ib}\gamma_b}$
&\multirow{2}{*}[-0.25cm]{$G_{12}^i$}
&$\begin{aligned}   &1+ \Phi_4\left(l_{ia},l_{ib},f_9(l_{ib})\right)- \Phi_4\left(l_{ia},l_{ib},0\right)\\
&-e^{-\frac{\theta_2}{ l_{ib}\gamma_b}}- \Phi_3\left(l_{ia},l_{ib},f_9(l_{ib})\right)+ \Phi_3\left(l_{ia},l_{ib},\frac{\theta_2}{l_{ib}\gamma_b}\right)
\end{aligned}$
\\ \cline{2-3} \cline{5-5}
& $1 - \theta_1 \theta_2 < 0$
& 
&
&$1- \Phi_4\left(l_{ia},l_{ib},0\right)-e^{-\frac{\theta_2}{ l_{ib}P_2}}+ \Phi_3\left(l_{ia},l_{ib},\frac{\theta_2}{l_{ib}\gamma_b}\right)$
\\ \hline 
&
&
&$G_{1}^i$
&$G_{11}^i + G_{12}^i$
\\ \hline \hline
\end{tabular}
    \vspace{-3mm}
\end{table*}

\begin{table*}
    \caption{Expressions for $G_2$}
    \vspace{-3mm}
\label{case G2}
\centering
\setlength{\tabcolsep}{4pt}
\begin{tabular}{c|c|c|c|c}
\hline \hline
\multicolumn{3}{c|}{Conditions} 
& Term
& Expression
\\
\hline \hline
$F_1<0$
&$\theta_{11}-\theta_1(\alpha-\theta_{11}(1-\alpha))<0$
&$\frac{a}{1-a} > \theta_{11}$
&\multirow{2}{*}[-0.1cm]{$G_{21}^i$} 
&$\begin{aligned}
    & \Phi_4\left(l_{ia},l_{ib},f_1(l_{ib})\right)- \Phi_4\left(l_{ia},l_{ib},0\right)\\ &- \Phi_1\left(l_{ia},l_{ib},f_1(l_{ib})\right) + \Phi_1\left(l_{ia},l_{ib},0\right)
\end{aligned}$
\\ \cline{1-3} \cline{5-5}
$F_1>0$
&$\theta_{11}-\theta_1(\alpha-\theta_{11}(1-\alpha))<0$
&$\frac{a}{1-a} > \theta_{11}$
&
&$ \Phi_1\left(l_{ia},l_{ib},0\right)- \Phi_4\left(l_{ia},l_{ib},0\right)$
\\ \hline
$F_1<0$
&$1-\theta_1\theta_2(1-\alpha)>0$
&$\begin{aligned}
   & \frac{a}{1-a} > \theta_{11} \ \cap \ \\
   &f_4(l_{ib})>f_1(l_{ib})
\end{aligned}$
&\multirow{2}{*}[-0.1cm]{$G_{22}^i$}
&$\begin{aligned}
    & \Phi_4\left(l_{ia},l_{ib},f_4(l_{ib})\right)- \Phi_4\left(l_{ia},l_{ib},f_1(l_{ib})\right)\\ &- \Phi_1\left(l_{ia},l_{ib},f_4(l_{ib})\right) + \Phi_1\left(l_{ia},l_{ib},f_1(l_{ib})\right)
\end{aligned}$
\\ \cline{1-3} \cline{5-5}
$F_1<0$
&$1-\theta_1\theta_2(1-\alpha)<0$
&$\frac{a}{1-a} > \theta_{11}$
& &$ \Phi_2\left(l_{ia},l_{ib},f_1(l_{ib})\right)- \Phi_4\left(l_{ia},l_{ib},f_1(l_{ib})\right)$
\\ \hline 
&
&
&$G_{2}^i$
&$G_{21}^i + G_{22}^i$
\\ \hline \hline
\end{tabular}
    \vspace{-3mm}
\end{table*}

\begin{table*}
    \caption{Expressions for $G_3$}
    \vspace{-3mm}
\label{case G3}
\centering
\setlength{\tabcolsep}{4pt}
\begin{tabular}{c|c|c|c}
\hline \hline
\multicolumn{2}{c|}{Conditions}
& Term 
& Expression
\\
\hline \hline
$F_1>0$
&$\begin{aligned}&\frac{a}{1-a} > \theta_{11} \ \cap \ \\
&\theta_{11}-\theta_1(\alpha-\theta_{11}(1-\alpha))<0\end{aligned}$
&\multirow{2}{*}[-0.35cm]{$G_{31}^i$}
&$\begin{aligned}
    &1- \Phi_1\left(l_{ia},l_{ib},0\right)-e^{-\frac{\theta_2}{ l_{ib}\gamma_b}} + \Phi_2\left(l_{ia},l_{ib},\frac{\theta_2}{ l_{ib}\gamma_b}\right)
\end{aligned}$
\\ \cline{1-2} \cline{4-4}
$F_1<0$
&$\begin{aligned}&\frac{a}{1-a} > \theta_{11} \ \cap \ \\
&\theta_{11}-\theta_1(\alpha-\theta_{11}(1-\alpha))<0\end{aligned}$
&
&$\begin{aligned}
    &1+ \Phi_1\left(l_{ia},l_{ib},f_1(l_{ib})\right)- \Phi_1\left(l_{ia},l_{ib},0\right)\\ &-e^{-\frac{\theta_2}{ l_{ib}\gamma_b}}- \Phi_2\left(l_{ia},l_{ib},f_1(l_{ib})\right) + \Phi_2\left(l_{ia},l_{ib},\frac{\theta_2}{ l_{ib}\gamma_b}\right)
\end{aligned}$
\\ \hline
$1-\theta_1\theta_2(1-\alpha)<0$
&$\begin{aligned}&\frac{a}{1-a} < \theta_{11} \ \cup \ \\
&\theta_{11}-\theta_1(\alpha-\theta_{11}(1-\alpha))>0\end{aligned}$
&\multirow{2}{*}[-0.35cm]{$G_{32}^i$}
&$\begin{aligned}
    &1- \Phi_4\left(l_{ia},l_{ib},0\right)-e^{-\frac{\theta_2}{ l_{ib}\gamma_b}} + \Phi_2\left(l_{ia},l_{ib},\frac{\theta_2}{ l_{ib}\gamma_b}\right)
\end{aligned}$
\\ \cline{1-2} \cline{4-4}
$1-\theta_1\theta_2(1-\alpha)>0$
&$\begin{aligned}&\frac{a}{1-a} < \theta_{11} \ \cup \ \\
&\theta_{11}-\theta_1(\alpha-\theta_{11}(1-\alpha))>0\end{aligned}$
&
&$\begin{aligned}
    &1+ \Phi_4\left(l_{ia},l_{ib},f_4(l_{ib})\right)- \Phi_4\left(l_{ia},l_{ib},0\right)\\& -e^{-\frac{\theta_2}{ l_{ib}\gamma_b}}- \Phi_2\left(l_{ia},l_{ib},f_4(l_{ib})\right) + \Phi_2\left(l_{ia},l_{ib},\frac{\theta_2}{ l_{ib}\gamma_b}\right)
\end{aligned}$
\\ \hline 
&
&$G_{3}^i$
&$G_{31}^i + G_{32}^i$
\\ \hline \hline
\end{tabular}
    \vspace{-3mm}
\end{table*}

\begin{table*}
    \caption{Expressions for $G_4$}
    \vspace{-3mm}
    \label{case G4}
\centering
\setlength{\tabcolsep}{4pt}
\resizebox{\textwidth}{!}{
\begin{tabular}{c|c|c|c|c}
\hline \hline
\multicolumn{3}{c|}{Conditions} 
& Term
& Expression
\\
\hline \hline
$F_2 > 0$
&$
\begin{aligned}
    &1-\theta_1\theta_2(1-\alpha)>0 \ \cap \\
    &\theta_{11}-\theta_1\alpha\\&+\theta_1\theta_{11}(1-\alpha))<0
\end{aligned}
$
&$f_5(l_{ib})>f_4(l_{ib})$
&\multirow{2}{*}[-0.4cm]{$G_{41}^i$}
& $\begin{aligned}& \Phi_4\left(l_{ia},l_{ib},f_5(l_{ib})\right)- \Phi_4\left(l_{ia},l_{ib},f_4(l_{ib})\right)\\&+  \Phi_1\left(l_{ia},l_{ib},f_4(l_{ib})\right)- \Phi_1\left(l_{ia},l_{ib},f_5(l_{ib})\right)\end{aligned}$
\\ \cline{1-3} \cline{5-5}
$F_2 < 0$
&$
\begin{aligned}
    &1-\theta_1\theta_2(1-\alpha)>0 \ \cap \\
    &\theta_{11}-\theta_1\alpha\\&+\theta_1\theta_{11}(1-\alpha))<0
\end{aligned}
$
&$\begin{aligned}
    &\min\left(f_5(l_{ib}),f_6(l_{ib})\right) >\\
    &f_4(l_{ib})
\end{aligned}$
&
&$\begin{aligned}
& \Phi_4\left(l_{ia},l_{ib},\min\left(f_5(l_{ib}),f_6(l_{ib})\right)\right) -  \Phi_4\left(l_{ia},l_{ib},f_4(l_{ib})\right)\\&- \Phi_1\left(l_{ia},l_{ib},\min\left(f_5(l_{ib}),f_6(l_{ib})\right)\right)+  \Phi_1\left(l_{ia},l_{ib},f_4(l_{ib})\right)
\end{aligned}$
\\ \hline
$\begin{aligned}
& F_2 > 0 \\
& \cap \ F_1 < 0
\end{aligned}$
&$1-\theta_1\theta_2(1-\alpha)<0$
&$f_3(l_{ib}) >f_1(l_{ib})
$
&\multirow{4}{*}[-0.4cm]{$G_{42}^i$}
&$\begin{aligned}
& \Phi_2\left(l_{ia},l_{ib},f_3(l_{ib})\right) -  \Phi_2\left(l_{ia},l_{ib},f_1(l_{ib})\right)\\&- \Phi_1\left(l_{ia},l_{ib},f_3(l_{ib})\right)+  \Phi_1\left(l_{ia},l_{ib},f_1(l_{ib})\right)
\end{aligned}$
\\ \cline{1-3} \cline{5-5}
$ \begin{aligned}
&F_2 > 0 
\\ & \cap \ F_1 < 0
\end{aligned}$
&$1-\theta_1\theta_2(1-\alpha)>0$
&$\begin{aligned}
    &\min\left(f_3(l_{ib}),f_4(l_{ib})\right) >\\
    &f_1(l_{ib})
\end{aligned}$
&
&$\begin{aligned}
& \Phi_2\left(l_{ia},l_{ib},\min\left(f_3(l_{ib}),f_4(l_{ib})\right)\right) -  \Phi_2\left(l_{ia},l_{ib},f_1(l_{ib})\right)\\&- \Phi_1\left(l_{ia},l_{ib},\min\left(f_3(l_{ib}),f_4(l_{ib})\right)\right)+  \Phi_1\left(l_{ia},l_{ib},f_1(l_{ib})\right)
\end{aligned}$
\\ \cline{1-3} \cline{5-5}
$\begin{aligned}
&F_2 < 0
\\ & \cap \ F_1 < 0
\end{aligned}$
&$1-\theta_1\theta_2(1-\alpha)<0$
&$\begin{aligned}
    &\min\left(f_3(l_{ib}),f_6(l_{ib})\right) >\\
    &f_1(l_{ib})
\end{aligned}$
&
&$\begin{aligned}
& \Phi_2\left(l_{ia},l_{ib},\min\left(f_3(l_{ib}),f_6(l_{ib})\right)\right) -  \Phi_2\left(l_{ia},l_{ib},f_1(l_{ib})\right)\\&- \Phi_1\left(l_{ia},l_{ib},\min\left(f_3(l_{ib}),f_6(l_{ib})\right)\right)+  \Phi_1\left(l_{ia},l_{ib},f_1(l_{ib})\right)
\end{aligned}$
\\ \cline{1-3} \cline{5-5}
$\begin{aligned}
& F_2 < 0 \\
& \cap \ F_1 < 0
\end{aligned}$
&$1-\theta_1\theta_2(1-\alpha)>0$
&$\begin{aligned}&\min\left(f_3(l_{ib}),f_4(l_{ib}), \right.\\& \left. f_6(l_{ib})\right)>f_1(l_{ib})\end{aligned}$
&
&$\begin{aligned}
& \Phi_2\left(l_{ia},l_{ib},\min\left(f_3(l_{ib}),f_4(l_{ib}),f_6(l_{ib})\right)\right) -  \Phi_2\left(l_{ia},l_{ib},f_1(l_{ib})\right)\\&- \Phi_1\left(l_{ia},l_{ib},\min\left(f_3(l_{ib}),f_4(l_{ib}),f_6(l_{ib})\right)\right)+  \Phi_1\left(l_{ia},l_{ib},f_1(l_{ib})\right)
\end{aligned}$
\\ \hline
$F_2 > 0$
&$1-\theta_1\theta_2(1-\alpha)>0$
&$\begin{aligned}
    &f_7(l_{ib}) >\\
    &\max\left(f_4(l_{ib}),f_5(l_{ib})\right)
\end{aligned}$
&\multirow{2}{*}[-0.35cm]{$G_{43}^i$}
&$\begin{aligned}
&e^{-\frac{\theta_{12}}{(1-\alpha) l_{ia}\gamma_a}}\left(e^{-f_7(l_{ib})}-e^{-\max\left(f_4(l_{ib}),f_5(l_{ib})\right)}\right)\\&- \Phi_1\left(l_{ia},l_{ib},f_7(l_{ib})\right) + \Phi_1\left(l_{ia},l_{ib},\max\left(f_4(l_{ib}),f_5(l_{ib})\right)\right)
\end{aligned}$
\\ \cline{1-3} \cline{5-5}
$F_2 < 0$
&$1-\theta_1\theta_2(1-\alpha)>0$
&$\begin{aligned}
    &\min\left(f_6(l_{ib}),f_7(l_{ib})\right) >\\
    &\max\left(f_4(l_{ib}),f_5(l_{ib})\right)
\end{aligned}$
&
&$\begin{aligned}
&e^{-\frac{\theta_{12}}{(1-\alpha) l_{ia}\gamma_a}}\left(e^{-\min\left(f_6(l_{ib}),f_7(l_{ib})\right)}-e^{-\max\left(f_4(l_{ib}),f_5(l_{ib})\right)}\right)\\&- \Phi_1\left(l_{ia},l_{ib},\min\left(f_6(l_{ib}),f_7(l_{ib})\right)\right) + \Phi_1\left(l_{ia},l_{ib},\max\left(f_4(l_{ib}),f_5(l_{ib})\right)\right)
\end{aligned}$
\\ \hline
$F_2 > 0$
&$1-\theta_1\theta_2(1-\alpha)<0$
&$f_7(l_{ib})>f_3(l_{ib})$
&\multirow{4}{*}[-8mm]{$G_{44}^i$}
&$\begin{aligned}
&e^{-\frac{\theta_{12}}{(1-\alpha) l_{ia}\gamma_a}}\left(e^{-f_7(l_{ib})}-e^{-f_3(l_{ib})}\right)\\&- \Phi_1\left(l_{ia},l_{ib},f_7(l_{ib})\right) + \Phi_1\left(l_{ia},l_{ib},f_3(l_{ib})\right)
\end{aligned}$
\\ \cline{1-3} \cline{5-5}
$F_2 > 0$
&$1-\theta_1\theta_2(1-\alpha)>0$
&$\begin{aligned} &\min\left(f_4(l_{ib}),f_7(l_{ib})\right)>\\&f_3(l_{ib})\end{aligned}$
&
&$\begin{aligned}
&e^{-\frac{\theta_{12}}{(1-\alpha) l_{ia}\gamma_a}}\left(e^{-\min\left(f_4(l_{ib}),f_7(l_{ib})\right)}-e^{-f_3(l_{ib})}\right)\\&-\left( \Phi_1\left(l_{ia},l_{ib},\min\left(f_4(l_{ib}),f_7(l_{ib})\right)\right) -  \Phi_1\left(l_{ia},l_{ib},f_3(l_{ib})\right)\right)
\end{aligned}$
\\ \cline{1-3} \cline{5-5}
$F_2 < 0$
&$1-\theta_1\theta_2(1-\alpha)<0$
&$\begin{aligned}&\min\left(f_6(l_{ib}),f_7(l_{ib})\right)> \\&f_3(l_{ib})\end{aligned}$
&
&$\begin{aligned}
&e^{-\frac{\theta_{12}}{(1-\alpha) l_{ia}\gamma_a}}\left(e^{-\min\left(f_6(l_{ib}),f_7(l_{ib})\right)}-e^{-f_3(l_{ib})}\right)\\&-\left( \Phi_2\left(l_{ia},l_{ib},\min\left(f_6(l_{ib}),f_7(l_{ib})\right)\right) -  \Phi_2\left(l_{ia},l_{ib},f_3(l_{ib})\right)\right)
\end{aligned}$
\\ \cline{1-3} \cline{5-5}
$F_2 < 0$
&$1-\theta_1\theta_2(1-\alpha)>0$
&$\begin{aligned}&\min\left(f_4(l_{ib}),f_6(l_{ib}), \right.\\& \left. f_7(l_{ib})\right)>f_3(l_{ib})\end{aligned}$
&
&$\begin{aligned}
&e^{-\frac{\theta_{12}}{(1-\alpha) l_{ia}\gamma_a}}\left(e^{-\min\left(f_4(l_{ib}),f_6(l_{ib}),f_7(l_{ib})\right)}-e^{-f_3(l_{ib})}\right)\\&-\left( \Phi_1\left(l_{ia},l_{ib},\min\left(f_4(l_{ib}),f_6(l_{ib}),f_7(l_{ib})\right)\right) -  \Phi_1\left(l_{ia},l_{ib},f_3(l_{ib})\right)\right)
\end{aligned}$
\\ \hline
$F_2 < 0$
&$\begin{aligned}
&1-\theta_1\theta_2(1-\alpha)>0 \\ 
&\cap \ 1-\theta_1\theta_2<0
\end{aligned}$
&$\begin{aligned}
    &f_5(l_{ib}) > \\&
    \max\left(f_4(l_{ib}),f_6(l_{ib})\right)
\end{aligned}$
&\multirow{2}{*}[-0.15cm]{$G_{45}^i$}
&$\begin{aligned}
& \Phi_4\left(l_{ia},l_{ib},f_5(l_{ib})\right)- \Phi_4\left(l_{ia},l_{ib},\max\left(f_4(l_{ib}),f_6(l_{ib})\right)\right)\\&- \Phi_3\left(l_{ia},l_{ib},f_5(l_{ib})\right) + \Phi_3\left(l_{ia},l_{ib},\max\left(f_4(l_{ib}),f_6(l_{ib})\right)\right)
\end{aligned}$
\\ \cline{1-3} \cline{5-5}
$F_2 < 0$
&$\begin{aligned}
&1-\theta_1\theta_2(1-\alpha)>0 \\ 
& \cap \ 1-\theta_1\theta_2>0
\end{aligned}$
&$\begin{aligned}
    &\min\left(f_5(l_{ib}),f_9(l_{ib})\right) > \\&
    \max\left(f_4(l_{ib}),f_6(l_{ib})\right)
\end{aligned}$
&
&$\begin{aligned}
& \Phi_4\left(l_{ia},l_{ib},\min\left(f_5(l_{ib}),f_9(l_{ib})\right)\right)- \Phi_4\left(l_{ia},l_{ib},\max\left(f_4(l_{ib}),f_6(l_{ib})\right)\right)\\&- \Phi_3\left(l_{ia},l_{ib},\min\left(f_5(l_{ib}),f_9(l_{ib})\right)\right) + \Phi_3\left(l_{ia},l_{ib},\max\left(f_4(l_{ib}),f_6(l_{ib})\right)\right)
\end{aligned}$
\\ \hline
$F_2 < 0$
&$1-\theta_1\theta_2(1-\alpha)<0$
&$f_3(l_{ib})>f_6(l_{ib})$
&\multirow{2}{*}[-0.15cm]{$G_{46}^i$}
&$\begin{aligned}
& \Phi_2\left(l_{ia},l_{ib},f_3(l_{ib})\right)- \Phi_2\left(l_{ia},l_{ib},f_6(l_{ib})\right)\\&- \Phi_3\left(l_{ia},l_{ib},f_3(l_{ib})\right) + \Phi_3\left(l_{ia},l_{ib},f_6(l_{ib})\right)
\end{aligned}$
\\ \cline{1-3} \cline{5-5}
$F_2 < 0$
&$1-\theta_1\theta_2(1-\alpha)>0$
&$\begin{aligned}&\min\left(f_3(l_{ib}),f_4(l_{ib})\right)> \\&f_6(l_{ib})\end{aligned}$
&
&$\begin{aligned}
& \Phi_2\left(l_{ia},l_{ib},\min\left(f_3(l_{ib}),f_4(l_{ib})\right)\right)- \Phi_2\left(l_{ia},l_{ib},f_6(l_{ib})\right)\\&- \Phi_3\left(l_{ia},l_{ib},\min\left(f_3(l_{ib}),f_4(l_{ib})\right)\right) + \Phi_3\left(l_{ia},l_{ib},f_6(l_{ib})\right)
\end{aligned}$
\\ \hline
$F_2 < 0$
&$1-\theta_1\theta_2(1-\alpha)>0$
&$\begin{aligned}
    f_8(l_{ib}) &> \max\left(f_4(l_{ib}), \right. \\&
    \left. f_5(l_{ib}),f_6(l_{ib})\right)
\end{aligned}$
&$G_{47}^i$
&$\begin{aligned}
&e^{-\frac{\theta_{12}}{(1-\alpha) l_{ia}\gamma_a}}\left(e^{-f_8(l_{ib})}-e^{-\max\left(f_4(l_{ib}),f_5(l_{ib}),f_6(l_{ib})\right)}\right)\\&-\left( \Phi_4\left(l_{ia},l_{ib},f_8(l_{ib})\right)- \Phi_4\left(l_{ia},l_{ib},\max\left(f_4(l_{ib}),f_5(l_{ib}),f_6(l_{ib})\right)\right)\right)
\end{aligned}$
\\ \hline
$F_2 < 0$
&$1-\theta_1\theta_2(1-\alpha)<0$
&$\begin{aligned}
    &f_8(l_{ib}) >\\
    &\max\left(f_3(l_{ib}),f_6(l_{ib})\right)
\end{aligned}$
&\multirow{2}{*}[-0.25cm]{$G_{48}^i$}
&$\begin{aligned}
&e^{-\frac{\theta_{12}}{(1-\alpha) l_{ia}\gamma_a}}\left(e^{-f_8(l_{ib})}-e^{-\max\left(f_3(l_{ib}),f_6(l_{ib})\right)}\right)\\&-\left( \Phi_2\left(l_{ia},l_{ib},f_8(l_{ib})\right)- \Phi_2\left(l_{ia},l_{ib},\max\left(f_3(l_{ib}),f_6(l_{ib})\right)\right)\right)
\end{aligned}$
\\ \cline{1-3} \cline{5-5}
$F_2 < 0$
&$1-\theta_1\theta_2(1-\alpha)>0$
&$\begin{aligned}
    &\min\left(f_4(l_{ib}),f_8(l_{ib})\right) >\\
    &\max\left(f_3(l_{ib}),f_6(l_{ib})\right)
\end{aligned}$
&
&$\begin{aligned}
&e^{-\frac{\theta_{12}}{(1-\alpha) l_{ia}\gamma_a}}\left(e^{-\min\left(f_4(l_{ib}),f_8(l_{ib})\right)}-e^{-\max\left(f_3(l_{ib}),f_6(l_{ib})\right)}\right)\\&-\left(\Phi_2\left( ,l_{ia}, ,l_{ib}\min\left(f_4(l_{ib}),f_8(l_{ib})\right)\right)- \Phi_2\left(l_{ia},l_{ib},\max\left(f_3(l_{ib}),f_6(l_{ib})\right)\right)\right)
\end{aligned}$
\\ \hline
&
&
&$G_{4}^i$
&$G_{41}^i + G_{42}^i + G_{43}^i + G_{44}^i + G_{45}^i + G_{46}^i+ G_{47}^i + G_{48}^i$
\\ \hline \hline
\end{tabular}
}
    \vspace{-3mm}
\end{table*}

\subsection{OP of User $a$}
In the following theorem, the OP of the user performing message splitting is provided. In order for the individual terms provided in tables to have non-zero values, all the conditions mentioned in the previous columns of the same row must hold. In other words, if a single condition is not satisfied, the corresponding term is equal to $0$.  
\begin{theorem}
\label{th1}
    The OP of user $a$ is given by 
    \begin{equation}
    \label{P_a}
    \begin{aligned}
        P_a^o &= G_1^iG_1^j + G_2^iG_2^j + G_2^iG_3^j + G_3^iG_2^j + G_4^iG_4^j + G_4^iG_5^j \\ &+ G_5^iG_4^j + G_4^iG_6^j + G_4^iG_7^j + G_5^iG_6^j + G_6^iG_4^j + G_6^iG_5^j \\ &+ G_7^iG_4^j + G_8^iG_8^j + G_8^iG_9^j + G_9^iG_8^j.
    \end{aligned}
    \end{equation}
    The expressions for $G_1^i-G_7^i$ can be found in Tables \ref{case G1}-\ref{case G7},
    \begin{equation}
        \begin{aligned}
            G_8^i =\frac{ l_{ib}\gamma_be^{-\frac{\theta_2}{ l_{ib}\gamma_b}}}{ l_{ib}\gamma_b+ l_{ia}\gamma_a\theta_2}-\frac{ l_{ib}\gamma_be^{-\frac{\theta_2(1+\theta_1)}{ l_{ib}\gamma_b}-\frac{\theta_1}{ l_{ia}\gamma_a}}}{ l_{ib}\gamma_b+ l_{ia}\gamma_a\theta_2}
        \end{aligned}
    \end{equation}
    and
    \begin{equation}
    \begin{aligned}
        G_9^i &= 1-e^{-\frac{\theta_1}{ l_{ia}\gamma_a}}\left(1-e^{-\frac{\theta_2(\theta_1+1)}{ l_{ib}\gamma_b}}\right)-e^{-\frac{\theta_2}{ l_{ib}\gamma_b}}\\ &-\left(\frac{ l_{ia}\theta_2\gamma_ae^{-\frac{\theta_1\theta_2}{ l_{ia}\theta_2\gamma_a}-\frac{\theta_2(\theta_1+1)}{ l_{ib}\gamma_b}}}{ l_{ia}\theta_2\gamma_a+ l_{ib}\gamma_b}-\frac{ l_{ia}\theta_2\gamma_ae^{-\frac{\theta_2}{ l_{ib}\gamma_b}}}{ l_{ia}\theta_2\gamma_a+ l_{ib}\gamma_b}\right).
        \end{aligned}
    \end{equation}
    The expressions for $G_1^j -G_9^j$ can be derived by changing $i$ to $j$.
\end{theorem}
\begin{IEEEproof}
    A detailed proof is provided in Appendix \ref{Proof1}. 
\end{IEEEproof}

\begin{table*}
    \caption{Expressions for $G_5$}
    \vspace{-3mm}
\label{case G5}
\centering
\setlength{\tabcolsep}{4pt}
\begin{tabular}{c|c|c|c|c|c}
\hline \hline
\multicolumn{4}{c|}{Conditions}
& Term
& Expression
\\
\hline \hline
\multirow{7}{*}[-1.3cm]{\rotatebox[origin=c]{90}{\centering$\frac{\alpha}{1-\alpha}>\theta_{11}$}}
& $F_1 < 0$
&$\begin{aligned}&\theta_{11}-\theta_1\alpha\\&+\theta_1\theta_{11}(1-\alpha))<0\end{aligned}$
&$\min\left(f_1(l_{ib}),f_5(l_{ib})\right)>0$
&\multirow{2}{*}[-0.25cm]{$G_{51}^i$}
& $\begin{aligned}& \Phi_4\left(l_{ia},l_{ib},\min\left(f_1(l_{ib}),f_5(l_{ib})\right)\right)- \Phi_4\left(l_{ia},l_{ib},0\right)\\&+  \Phi_1\left(l_{ia},l_{ib},0\right)- \Phi_1\left(l_{ia},l_{ib},\min\left(f_1(l_{ib}),f_5(l_{ib})\right)\right)\end{aligned}$
\\ \cline{2-4} \cline{6-6}
& $F_1 > 0$
&$\begin{aligned}&\theta_{11}-\theta_1\alpha\\&+\theta_1\theta_{11}(1-\alpha))<0\end{aligned}$
&$f_5(l_{ib})>0$
&
& $\begin{aligned}& \Phi_4\left(l_{ia},l_{ib},f_5(l_{ib})\right)- \Phi_4\left(l_{ia},l_{ib},0\right)\\&+  \Phi_1\left(l_{ia},l_{ib},0\right)- \Phi_1\left(l_{ia},l_{ib},f_5(l_{ib})\right)\end{aligned}$
\\ \cline{2-6}
& $F_1 < 0$
&$1-\theta_1\theta_2<0$
&$f_5(l_{ib})>f_1(l_{ib})$
&\multirow{2}{*}[-0.25cm]{$G_{52}^i$}
& $\begin{aligned}& \Phi_4\left(l_{ia},l_{ib},f_5(l_{ib})\right)- \Phi_4\left(l_{ia},l_{ib},f_1(l_{ib})\right)\\&+  \Phi_1\left(l_{ia},l_{ib},f_1(l_{ib})\right)- \Phi_1\left(l_{ia},l_{ib},f_5(l_{ib})\right)\end{aligned}$
\\ \cline{2-4} \cline{6-6}
& $F_1 < 0$
&$1-\theta_1\theta_2<0$
&$\min\left(f_4(l_{ib}),f_5(l_{ib})\right)>f_1(l_{ib})$
&
& $\begin{aligned}& \Phi_4\left(l_{ia},l_{ib},\min\left(f_4(l_{ib}),f_5(l_{ib})\right)\right)- \Phi_4\left(l_{ia},l_{ib},f_1(l_{ib})\right)\\&+  \Phi_1\left(l_{ia},l_{ib},f_1(l_{ib})\right)- \Phi_1\left(l_{ia},l_{ib},\min\left(f_4(l_{ib}),f_5(l_{ib})\right)\right)\end{aligned}$
\\ \cline{2-6}
& $F_1 < 0$
&
&$\min\left(f_1(l_{ib}),f_7(l_{ib})\right)>f_5(l_{ib})$
&\multirow{2}{*}[-0.4cm]{$G_{53}^i$}
&$\begin{aligned}
&e^{-\frac{\theta_{12}}{(1-\alpha) l_{ia}\gamma_a}}\left(e^{-\min\left(f_1(l_{ib}),f_7(l_{ib})\right)}-e^{-f_5(l_{ib})}\right)\\&-\left( \Phi_2\left(l_{ia},l_{ib},\min\left(f_1(l_{ib}),f_7(l_{ib})\right)\right)- \Phi_2\left(l_{ia},l_{ib},f_5(l_{ib})\right)\right)
\end{aligned}$
\\ \cline{2-4} \cline{6-6}
& $F_1 > 0$
&
&$f_7(l_{ib})>f_5(l_{ib})$
&
&$\begin{aligned}
&e^{-\frac{\theta_{12}}{(1-\alpha) l_{ia}\gamma_a}}\left(e^{-f_7(l_{ib})}-e^{-f_5(l_{ib})}\right)\\&-\left( \Phi_2\left(l_{ia},l_{ib},f_7(l_{ib})\right)- \Phi_2\left(l_{ia},l_{ib},f_5(l_{ib})\right)\right)
\end{aligned}$
\\ \cline{2-6}
& $F_1 < 0$
&
&$f_3(l_{ib})>\max\left(f_1(l_{ib}),f_5(l_{ib})\right)$
&$G_{54}^i$
&$\begin{aligned}
&e^{-\frac{\theta_{12}}{(1-\alpha) l_{ia}\gamma_a}}\left(e^{-f_3(l_{ib})}-e^{-\max\left(f_1(l_{ib}),f_5(l_{ib})\right)}\right)\\&-\left( \Phi_2\left(l_{ia},l_{ib},f_3(l_{ib})\right)- \Phi_2\left(l_{ia},l_{ib},\max\left(f_1(l_{ib}),f_5(l_{ib})\right)\right)\right)
\end{aligned}$
\\ \hline
&
&
&
&$G_{5}^i$
&$G_{51}^i + G_{52}^i + G_{53}^i + G_{54}^i$
\\ \hline \hline
\end{tabular}
    \vspace{-3mm}
\end{table*}

\begin{table*}
\caption{Expressions for $G_6$}
    \vspace{-3mm}
\label{case G6}
\centering
\setlength{\tabcolsep}{4pt}
\begin{tabular}{c|c|c|c|c|c}
\hline \hline
\multicolumn{4}{c|}{Conditions} 
& Term
& Expression
\\
\hline \hline
\multirow{6}{*}{\rotatebox[origin=c]{90}{\centering $\begin{aligned} &\frac{\alpha}{1-\alpha}>\theta_{11}\ \cap \ \theta_{11} -\theta_1(\alpha-\theta_{11}(1-\alpha))<0\end{aligned}$}}
&\multirow{2}{*}[-2.5mm]{$F_1<0$}
&
&$\min\left(f_6(l_{ib}),f_7(l_{ib})\right)>f_1(l_{ib})$
&$G_{61}^i$
&$\begin{aligned} & \Phi_1\left(l_{ia},l_{ib},\min\left(f_6(l_{ib}),f_7(l_{ib})\right)\right)- \Phi_1\left(l_{ia},l_{ib},f_1(l_{ib})\right)\\&+ \Phi_3\left(l_{ia},l_{ib},f_1(l_{ib})\right)- \Phi_3\left(l_{ia},l_{ib},\min\left(f_6(l_{ib}),f_7(l_{ib})\right)\right)\end{aligned}$
\\ \cline{4-6}
&
&
&$f_8(l_{ib})>\max\left(f_1(l_{ib}),f_7(l_{ib})\right)$
&$G_{62}^i$
&$\begin{aligned} &e^{-\frac{\theta_{12}}{(1-\alpha) l_{ia}\gamma_a}}\left(e^{-f_8(l_{ib})}-e^{-\max\left(f_1(l_{ib}),f_7(l_{ib})\right)}\right)\\& - \Phi_3\left(l_{ia},l_{ib},f_8(l_{ib})\right)+ \Phi_3\left(l_{ia},l_{ib},\max\left(f_1(l_{ib}),f_7(l_{ib})\right)\right) \end{aligned}$
\\ \cline{2-6}
& $F_1<0$
&
&$\min\left(f_1(l_{ib}),f_4(l_{ib})\right)>\frac{\theta_2}{l_{ib}\gamma_b}$
&\multirow{2}{*}[-0.5cm]{$G_{63}^i$}
&$\begin{aligned} & \Phi_2\left(l_{ia},l_{ib},\min\left(f_1(l_{ib}),f_4(l_{ib})\right)\right)- \Phi_2\left(l_{ia},l_{ib},\frac{\theta_2}{l_{ib}\gamma_b}\right)\\&+ \Phi_3\left(l_{ia},l_{ib},\frac{\theta_2}{l_{ib}\gamma_b}\right)- \Phi_3\left(l_{ia},l_{ib},\min\left(f_1(l_{ib}),f_4(l_{ib})\right)\right)\end{aligned}$
\\ \cline{2-4} \cline{6-6}
& $F_1>0$
&
&$f_4(l_{ib})>\frac{\theta_2}{l_{ib}\gamma_b}$
&
&$\begin{aligned} & \Phi_2\left(l_{ia},l_{ib},f_4(l_{ib})\right)- \Phi_2\left(l_{ia},l_{ib},\frac{\theta_2}{l_{ib}\gamma_b}\right)\\&+ \Phi_3\left(l_{ia},l_{ib},\frac{\theta_2}{l_{ib}\gamma_b}\right)- \Phi_3\left(l_{ia},l_{ib},f_4(l_{ib})\right)\end{aligned}$
\\ \cline{2-6}
& $F_1<0$
&
&$\min\left(f_1(l_{ib}),f_8(l_{ib})\right)>f_3(l_{ib})$
&\multirow{2}{*}[-0.35cm]{$G_{64}^i$}
&$\begin{aligned} &e^{-\frac{\theta_{12}}{(1-\alpha) l_{ia}\gamma_a}}\left(e^{-\min\left(f_1(l_{ib}),f_8(l_{ib})\right)}-e^{-f_3(l_{ib})}\right)\\& - \Phi_3\left(l_{ia},l_{ib},\min\left(f_1(l_{ib}),f_8(l_{ib})\right)\right)+ \Phi_3\left(l_{ia},l_{ib},f_3(l_{ib})\right) \end{aligned}$
\\ \cline{2-4} \cline{6-6}
& $F_1>0$
&
&$f_8(l_{ib})>f_3(l_{ib})$
&
&$\begin{aligned} &e^{-\frac{\theta_{12}}{(1-\alpha) l_{ia}\gamma_a}}\left(e^{-f_8(l_{ib})}-e^{-f_3(l_{ib})}\right)\\& - \Phi_3\left(l_{ia},l_{ib},f_8(l_{ib})\right)+ \Phi_3\left(l_{ia},l_{ib},f_3(l_{ib})\right) 
\end{aligned}$
\\ \hline
\multirow{9}{*}[-2mm]{\rotatebox[origin=c]{90}{\centering $\begin{aligned} &\frac{\alpha}{1-\alpha}<\theta_{11} \ \cup \ \theta_{11} -\theta_1(\alpha-\theta_{11}(1-\alpha))>0\end{aligned}$}}
&\multirow{3}{*}[-5mm]{$F_3>0$}
& $1-\theta_1\theta_2>0$
&$\min\left(f_5(l_{ib}),f_9(l_{ib})\right)>f_4(l_{ib})$
&\multirow{2}{*}[-2mm]{$G_{65}^i$}
&$\begin{aligned} & \Phi_4\left(l_{ia},l_{ib},\min\left(f_5(l_{ib}),f_9(l_{ib})\right)\right)- \Phi_4\left(l_{ia},l_{ib},f_4(l_{ib})\right)\\&+ \Phi_3\left(l_{ia},l_{ib},f_4(l_{ib})\right)- \Phi_3\left(l_{ia},l_{ib},\min\left(f_5(l_{ib}),f_9(l_{ib})\right)\right)\end{aligned}$
\\ \cline{3-4} \cline{6-6}
&
& $1-\theta_1\theta_2<0$
&$f_5(l_{ib}) > f_4(l_{ib})$
&
&$\begin{aligned} & \Phi_4\left(l_{ia},l_{ib},f_5(l_{ib})\right)- \Phi_4\left(l_{ia},l_{ib},f_4(l_{ib})\right)\\&+ \Phi_3\left(l_{ia},l_{ib},f_4(l_{ib})\right)- \Phi_3\left(l_{ia},l_{ib},f_5(l_{ib})\right)\end{aligned}$
\\ \cline{3-6}
&
&
&$f_8(l_{ib}) > \max\left(f_4(l_{ib}),f_5(l_{ib})\right)$
&$G_{66}^i$
&$\begin{aligned} &e^{-\frac{\theta_{12}}{(1-\alpha) l_{ia}\gamma_a}}\left(e^{-f_8(l_{ib})}-e^{-\max\left(f_4(l_{ib}),f_5(l_{ib})\right)}\right)\\& - \Phi_3\left(l_{ia},l_{ib},f_8(l_{ib})\right)+ \Phi_3\left(l_{ia},l_{ib},\max\left(f_4(l_{ib}),f_5(l_{ib})\right)\right) 
\end{aligned}$
\\ \cline{2-6}
& $F_3>0$
&
&$\min\left(f_3(l_{ib}),f_4(l_{ib})\right) > \frac{\theta_2}{l_{ib}\gamma_b}$
&\multirow{2}{*}[-0.5cm]{$G_{67}^i$}
 &$\begin{aligned} & \Phi_2\left(l_{ia},l_{ib},\min\left(f_3(l_{ib}),f_4(l_{ib})\right)\right)- \Phi_4\left(l_{ia},l_{ib},\frac{\theta_2}{l_{ib}\gamma_b}\right)\\&+ \Phi_3\left(l_{ia},l_{ib},\frac{\theta_2}{l_{ib}\gamma_b}\right)- \Phi_3\left(l_{ia},l_{ib},\min\left(f_3(l_{ib}),f_4(l_{ib})\right)\right)\end{aligned}$
 \\ \cline{2-4} \cline{6-6}
 & $F_3<0$
 &
 &$f_3(l_{ib}) > \frac{\theta_2}{l_{ib}\gamma_b}$
 &
 &$\begin{aligned} & \Phi_2\left(l_{ia},l_{ib},f_3(l_{ib})\right)- \Phi_4\left(l_{ia},l_{ib},\frac{\theta_2}{l_{ib}\gamma_b}\right)\\&+ \Phi_3\left(l_{ia},l_{ib},\frac{\theta_2}{l_{ib}\gamma_b}\right)- \Phi_3\left(l_{ia},l_{ib},f_3(l_{ib})\right)\end{aligned}$
 \\ \cline{2-6}
 & $F_3>0$
 &
 &$\min\left(f_4(l_{ib}),f_8(l_{ib})\right) > f_3(l_{ib})$
 &\multirow{2}{*}[-0.3cm]{$G_{68}^i$}
 &$\begin{aligned} &e^{-\frac{\theta_{12}}{(1-\alpha) l_{ia}\gamma_a}}\left(e^{-\min\left(f_4(l_{ib}),f_8(l_{ib})\right)}-e^{-f_3(l_{ib})}\right)\\& - \Phi_3\left(l_{ia},l_{ib},\min\left(f_4(l_{ib}),f_8(l_{ib})\right)\right)+ \Phi_3\left(l_{ia},l_{ib},f_3(l_{ib})\right) \end{aligned}$
 \\ \cline{2-4} \cline{6-6}
 & $F_3<0$
 &
 &$f_8(l_{ib}) > f_3(l_{ib})$
 &
 &$\begin{aligned} &e^{-\frac{\theta_{12}}{(1-\alpha) l_{ia}\gamma_a}}\left(e^{-f_8(l_{ib})}-e^{-f_3(l_{ib})}\right)\\& - \Phi_3\left(l_{ia},l_{ib},f_8(l_{ib}\right)+ \Phi_3\left(l_{ia},l_{ib},f_3(l_{ib})\right) \end{aligned}$
 \\ \hline
 &
 &
 &
 &$G_{6}^i$
 &$G_{61}^i + G_{62}^i + G_{63}^i + G_{64}^i + G_{65}^i + G_{66}^i + G_{67}^i + G_{68}^i$
 \\ \hline \hline
\end{tabular}
    \vspace{-3mm}
\end{table*}

\begin{table*}
    \caption{Expressions for $G_7$}
    \vspace{-3mm}
\label{case G7}
\centering
\setlength{\tabcolsep}{4pt}
\begin{tabular}{c|c|c|c|c}
\hline \hline
\multicolumn{3}{c|}{Conditions} 
& Term
& Expression
\\
\hline \hline
\multirow{3}{*}{\rotatebox[origin=c]{90}{\centering $\begin{aligned} &\frac{\alpha}{1-\alpha}>\theta_{11}\ \cap \ \theta_{11} \\&-\theta_1(\alpha-\theta_{11}(1-\alpha))<0\end{aligned}$}}
& $F_1>0$
&$f_7(l_{ib}) > 0$
&\multirow{2}{*}[-0.5cm]{$G_{71}^i$}
&$\begin{aligned} &1-\exp\left(-\frac{\theta_2}{ l_{ib}\gamma_b}\right) + \Phi_1\left(l_{ia},l_{ib},f_7(l_{ib})\right)+ \Phi_1\left(l_{ia},l_{ib},0\right)\\& - \Phi_2\left(l_{ia},l_{ib},f_7(l_{ib})\right)+ \Phi_2\left(l_{ia},l_{ib},\frac{\theta_2}{l_{ib}\gamma_b}\right) \end{aligned}$
\\ \cline{2-3} \cline{5-5}
& $F_1<0$
&$\min\left(f_1(l_{ib}),f_7(l_{ib})\right) > 0$
&
&$\begin{aligned} &1-\exp\left(-\frac{\theta_2}{ l_{ib}\gamma_b}\right) + \Phi_1\left(l_{ia},l_{ib},\min\left(f_1(l_{ib}),f_7(l_{ib})\right)\right)+ \Phi_1\left(l_{ia},l_{ib},0\right)\\& - \Phi_2\left(l_{ia},l_{ib},\min\left(f_1(l_{ib}),f_7(l_{ib})\right)\right)+ \Phi_2\left(l_{ia},l_{ib},\frac{\theta_2}{l_{ib}\gamma_b}\right) \end{aligned}$
\\ \cline{2-5}
&
&$f_3(l_{ib}) > \max\left(f_7(l_{ib}),0\right)$
&$G_{72}^i$
&$\begin{aligned} &1-\exp\left(-\frac{\theta_2}{ l_{ib}\gamma_b}\right)+e^{-\frac{\theta_{12}}{(1-\alpha) l_{ia}\gamma_a}}\left(e^{-f_3(l_{ib})}-e^{-\max\left(f_7(l_{ib}),0\right)}\right)\\& - \Phi_2\left(l_{ia},l_{ib},f_3(l_{ib})\right)+ \Phi_2\left(l_{ia},l_{ib},\max\left(f_7(l_{ib}),0\right)\right) 
\end{aligned}$
\\ \hline
\multirow{3}{*}{\rotatebox[origin=c]{90}{\centering $\begin{aligned} &\frac{\alpha}{1-\alpha}<\theta_{11} \ \cup \ \theta_{11} \\&-\theta_1(\alpha-\theta_{11}(1-\alpha))>0\end{aligned}$}}
& $\begin{aligned}&1-\theta_1\theta_2\\&\times(1-\alpha)>0\end{aligned}$
&$\min\left(f_4(l_{ib}),f_5(l_{ib})\right) > 0$
&\multirow{2}{*}[-0.4cm]{$G_{73}^i$}
&$\begin{aligned} &1-\exp\left(-\frac{\theta_2}{ l_{ib}\gamma_b}\right)+ \Phi_4\left(l_{ia},l_{ib},\min\left(f_4(l_{ib}),f_5(l_{ib})\right)\right)+ \Phi_4\left(l_{ia},l_{ib},0\right)\\& - \Phi_2\left(l_{ia},l_{ib},\min\left(f_4(l_{ib}),f_5(l_{ib})\right)\right)+ \Phi_2\left(l_{ia},l_{ib},0\right) 
\end{aligned}$
\\ \cline{2-3} \cline{5-5}
& $\begin{aligned}&1-\theta_1\theta_2\\&\times(1-\alpha)<0\end{aligned}$
&$f_5(l_{ib}) > 0$
&
&$\begin{aligned} &1-\exp\left(-\frac{\theta_2}{ l_{ib}\gamma_b}\right)+ \Phi_4\left(l_{ia},l_{ib},f_5(l_{ib})\right)+ \Phi_4\left(l_{ia},l_{ib},0\right)\\& - \Phi_2\left(l_{ia},l_{ib},f_5(l_{ib})\right)+ \Phi_2\left(l_{ia},l_{ib},0\right) 
\end{aligned}$
\\ \cline{2-5}
&
&$f_3(l_{ib}) > \max\left(f_5(l_{ib}),\frac{\theta_2}{l_{ib}\gamma_b}\right)$
&$G_{74}^i$
&$\begin{aligned} &1-\exp\left(-\frac{\theta_2}{ l_{ib}\gamma_b}\right)+e^{-\frac{\theta_{12}}{(1-\alpha) l_{ia}\gamma_a}}\left(e^{-f_3(l_{ib})}-e^{-\max\left(f_5(l_{ib}),\frac{\theta_2}{l_{ib}\gamma_b}\right)}\right)\\& - \Phi_2\left(l_{ia},l_{ib},f_3(l_{ib})\right)+ \Phi_2\left(l_{ia},l_{ib},\max\left(f_5(l_{ib}),\frac{\theta_2}{l_{ib}\gamma_b}\right)\right) 
\end{aligned}$
\\ \hline
&
&
&$G_7^i$
&$G_{71}^i + G_{72}^i + G_{73}^i + G_{74}^i$
\\ \hline \hline
\end{tabular}
    \vspace{-3mm}
\end{table*}

\subsection{OP of User $b$}
In the following theorem, the OP of user $b$ is provided.
\begin{theorem}
\label{th2}
    The OP of user $b$ is given by
    \begin{equation} \label{Outage b}
        \begin{aligned}
            P_b^o &= G_1^iG_1^j + G_2^iG_2^j + G_2^iG_3^j + G_3^iG_2^j + G_{10}^iG_{10}^j\\ &+G_{10}^iG_{11}^j + G_{11}^iG_{10}^j,
        \end{aligned}
    \end{equation}
    where the expressions for $G_1^i-G_3^i$ can be found in Tables \ref{case G1}-\ref{case G3}, while
    \begin{equation}
\label{G_10}
\begin{aligned}
    G_{10}^i =  \Phi_4\left(l_{ia},l_{ib},0\right)- \Phi_4\left(l_{ia},l_{ib},\frac{\theta_2}{l_{ib}\gamma_b}\right),
    \end{aligned}
\end{equation}
    and
    \begin{equation}
\label{G_11}
\begin{aligned}
    G_{11}^i = 1 - e^{-\frac{\theta_2}{ l_{ib}\gamma_b}}-  \Phi_4\left(l_{ia},l_{ib},0\right)+ \Phi_4\left(l_{ia},l_{ib},\frac{\theta_2}{l_{ib}\gamma_b}\right).
    \end{aligned}
\end{equation}
    The expressions for $G_{10}^j$ and $G_{11}^j$ can be derived be interchanging $i$ with $j$ in \eqref{G_10} and \eqref{G_11}, respectively.

\end{theorem}

\begin{IEEEproof}
    The proof is provided in Appendix \ref{Proof2}.
\end{IEEEproof}

\section{Numerical Results and Simulations}
In this section, the performance of the considered network is illustrated and the theoretical analysis are corroborated by Monte Carlo simulations. We assume that the path loss factor is given by $l_{km} = cd_{km}^{-n}$ with $c = 10^{-3}$ and $n = 2.5$. 

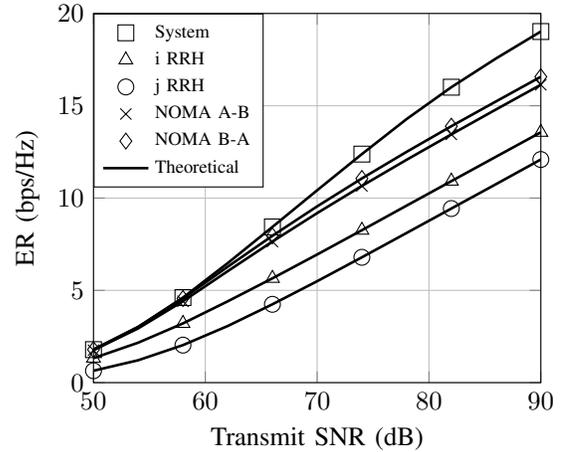
\begin{figure}
\centering
\begin{tikzpicture}
   \begin{axis}[
   width = 0.85\linewidth,
   xlabel = {Transmit SNR (dB)},
   ylabel = {ER (bps/Hz)},
   ymin = 0,
   ymax = 20,
   xmin = 50,
   xmax = 90,
   grid = major,
   legend entries = {{System},{i RRH},{j RRH},{NOMA A-B},{NOMA B-A},{Theoretical}},
   legend cell align = {left},
   legend style = {font = \scriptsize},
   legend style={at={(0,1)},anchor=north west}
   ]
   \addplot[
    black,
    mark = square,
    mark repeat = 2,
    mark size = 3,
    mark phase = 0,
    only marks,
    ]
    table {Data/ErgodicRate/ErgodicvsSNR/SumErgodicRSMA.dat};   
    \addplot[
    black,
    mark = triangle,
    mark repeat = 2,
    mark size = 3,
    mark phase = 0,
    only marks,
    ]
    table {Data/ErgodicRate/ErgodicvsSNR/SumErgodicRSMAIRRH.dat};
    \addplot[
    black,
    mark = o,
    mark repeat = 2,
    mark size = 3,
    mark phase = 0,
    only marks,
    ]
    table {Data/ErgodicRate/ErgodicvsSNR/SumErgodicRSMAJRRH.dat};
    \addplot[
    black,
    mark = x,
    mark repeat = 2,
    mark size = 3,
    mark phase = 0,
    only marks,
    ]
    table {Data/ErgodicRate/ErgodicvsSNR/SumErgodicNOMA1.dat};
    \addplot[
    black,
    mark = diamond,
    mark repeat = 2,
    mark size = 3,
    mark phase = 0,
    only marks,
    ]
    table {Data/ErgodicRate/ErgodicvsSNR/SumErgodicNOMA2.dat}; 
    \addplot[
    black,
    mark = square,
    mark repeat = 10,
    mark size = 3,
    mark phase = 0,
    no marks,
    line width = 1pt
    ]
    table {Data/ErgodicRate/ErgodicvsSNR/SumErgodicRSMA.dat};   
    \addplot[
    black,
    mark = triangle,
    mark repeat = 10,
    mark size = 3,
    mark phase = 0,
    no marks,
    line width = 1pt
    ]
    table {Data/ErgodicRate/ErgodicvsSNR/SumErgodicRSMAIRRH.dat};
    \addplot[
    black,
    mark = o,
    mark repeat = 10,
    mark size = 3,
    mark phase = 0,
    no marks,
    line width = 1pt
    ]
    table {Data/ErgodicRate/ErgodicvsSNR/SumErgodicRSMAJRRH.dat};     
    \addplot[
    black,
    mark = triangle,
    mark repeat = 10,
    mark size = 3,
    mark phase = 0,
    no marks,
    line width = 1pt
    ]
    table {Data/ErgodicRate/ErgodicvsSNR/SumErgodicNOMA1.dat};   
    \addplot[
    black,
    mark = o,
    mark repeat = 10,
    mark size = 3,
    mark phase = 0,
    no marks,
    line width = 1pt
    ]
    table {Data/ErgodicRate/ErgodicvsSNR/SumErgodicNOMA2.dat};
   \end{axis}
\end{tikzpicture}
\caption{ER vs transmit SNR.}
   \vspace{-2mm}
\label{ER vs SNR RSMA}
\end{figure}

In Fig. \ref{ER vs SNR RSMA}, the ER is plotted against the transmit SNR for $d_{ia} = 5$ m, $d_{ib} = 15$ m, $d_{ja} = 17$ m, and $d_{jb} = 8$ m. Specifically, this figure includes the ER of the system when the messages are decoded by the optimal RRH, the ER when a particular RRH is available for transmission regardless of the channel between the user and each RRH, and the ER of the two decoding orders of NOMA. The results in this figure are extracted for the optimal value of $\alpha$, which maximizes the ER of the system, for each SNR. As expected, increasing the SNR leads to a higher ER in every case, since the SINR increases in \eqref{Ergodic Split} and \eqref{Ergodic User B}. The better performance of RRH $i$ over RRH $j$ can be explained by taking into account the decoding order and the distance of each user from the RRHs. 
While the proposed scheme is slightly better than those implementing NOMA for lower SNRs, the improvement becomes significant for SNR $>75$ dB, with tendencies to become even greater considering the slope of each line. When implementing NOMA, decoding the message of user $b$ initially leads to higher ERs, which can be explained by considering the distances between users and RRHs. In particular, it is preferable to decode the user with the better SINR second, so that it can contribute to the system ER without interference. Considering that user $a$ is closer to the RRHs, it is optimal to decode user $a$ last. 

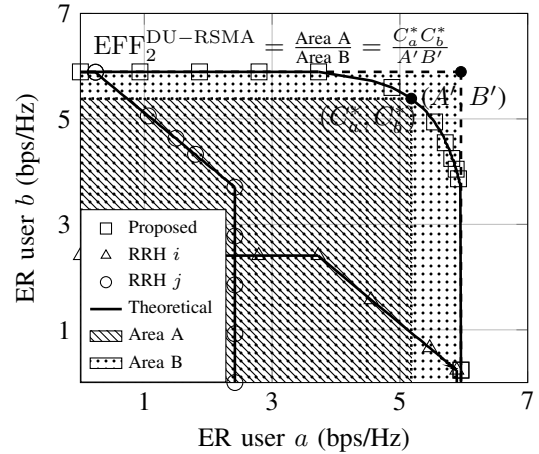
\begin{figure}
\centering
\begin{tikzpicture}
   \begin{axis}[
   width = 0.85\linewidth,
   xlabel = {ER user $a$ (bps/Hz)},
   ylabel = {ER user $b$ (bps/Hz)},
   ymin = 0,
   ymax = 7,
   xmin = 0,
   xmax = 7,
   ytick = {1, 3, 5, 7},
   yticklabels = {1, 3, 5, 7},
   xtick = {1, 3, 5, 7},
   xticklabels = {1, 3, 5, 7},
   grid = major,
   legend entries = {{Proposed},{},{},,{RRH $i$},{},{},{RRH $j$},{},{},{Theoretical},{},{},{},{},{},{},{},{},{},{},{},{},{Area A},{Area B}},
   legend cell align = {left},
   legend style = {font = \scriptsize},
   legend style={at={(0,0)},anchor=south west},
   legend image post style={scale=0.7},
   ]
   \addplot[
    black,
    mark = square,
    mark repeat = 8,
    mark size = 3,
    only marks,
    ]
    table {Data/ErgodicRate/CapacityRegion/CapacityFF/SumErgodic.dat};   
    \addplot[
    black,
    mark = square,
    mark repeat = 1,
    mark size = 3,
    only marks,
    ]
    table {Data/ErgodicRate/CapacityRegion/CapacityFF/x1.dat}; \addplot[
    black,
    mark = square,
    mark repeat = 5,
    mark size = 3,
    only marks,
    ]
    table {Data/ErgodicRate/CapacityRegion/CapacityFF/y1.dat};    
    \addplot[
    black,
    mark = triangle,
    mark repeat = 8,
    mark size = 3,
    mark phase = 42,
    only marks,
    ]
    table {Data/ErgodicRate/CapacityRegion/CapacityFF/IRRH.dat};
    \addplot[
    black,
    mark = triangle,
    mark repeat = 1,
    mark size = 3,
    only marks,
    ]
    table {Data/ErgodicRate/CapacityRegion/CapacityFF/x3.dat};
    \addplot[
    black,
    mark = triangle,
    mark repeat = 5,
    mark size = 3,
    only marks,
    ]
    table {Data/ErgodicRate/CapacityRegion/CapacityFF/y3.dat};
    \addplot[
    black,
    mark = o,
    mark repeat = 8,
    mark size = 3,
    mark phase = 27,
    only marks,
    ]
    table {Data/ErgodicRate/CapacityRegion/CapacityFF/JRRH.dat};
    \addplot[
    black,
    mark = o,
    mark repeat = 5,
    mark size = 3,
    only marks,
    ]
    table {Data/ErgodicRate/CapacityRegion/CapacityFF/x4.dat};
    \addplot[
    black,
    mark = o,
    mark repeat = 1,
    mark size = 3,
    only marks,
    ]
    table {Data/ErgodicRate/CapacityRegion/CapacityFF/y4.dat};
    \addplot[
    black,
    mark = square,
    mark repeat = 6,
    mark size = 3,
    mark phase = 4,
    no marks,
    line width = 1pt
    ]
    table {Data/ErgodicRate/CapacityRegion/CapacityFF/SumErgodic.dat};  
    \addplot[
    black,
    mark = square,
    mark repeat = 6,
    mark size = 3,
    mark phase = 4,
    no marks,
    line width = 1pt
    ]
    table {Data/ErgodicRate/CapacityRegion/CapacityFF/x1.dat};
    \addplot[
    black,
    mark = square,
    mark repeat = 6,
    mark size = 3,
    mark phase = 4,
    no marks,
    line width = 1pt
    ]
    table {Data/ErgodicRate/CapacityRegion/CapacityFF/y1.dat};
    \addplot[
    black,
    mark = triangle,
    mark repeat = 6,
    mark size = 3,
    mark phase = 4,
    no marks,
    line width = 1pt
    ]
    table {Data/ErgodicRate/CapacityRegion/CapacityFF/IRRH.dat};
    \addplot[
    black,
    mark = triangle,
    mark repeat = 6,
    mark size = 3,
    mark phase = 4,
    no marks,
    line width = 1pt
    ]
    table {Data/ErgodicRate/CapacityRegion/CapacityFF/x3.dat};
    \addplot[
    black,
    mark = triangle,
    mark repeat = 6,
    mark size = 3,
    mark phase = 4,
    no marks,
    line width = 1pt
    ]
    table {Data/ErgodicRate/CapacityRegion/CapacityFF/y3.dat};
    \addplot[
    black,
    mark = o,
    mark repeat = 2,
    mark size = 3,
    mark phase = 0,
    no marks,
    line width = 1pt
    ]
    table {Data/ErgodicRate/CapacityRegion/CapacityFF/JRRH.dat};
    \addplot[
    black,
    mark = o,
    mark repeat = 2,
    mark size = 3,
    mark phase = 0,
    no marks,
    line width = 1pt
    ]
    table {Data/ErgodicRate/CapacityRegion/CapacityFF/x4.dat};
    \addplot[
    black,
    mark = o,
    mark repeat = 2,
    mark size = 3,
    mark phase = 0,
    no marks,
    line width = 1pt
    ]
    table {Data/ErgodicRate/CapacityRegion/CapacityFF/y4.dat};
    \addplot[
    black,
    mark = o,
    mark repeat = 2,
    mark size = 3,
    mark phase = 0,
    no marks,
    style = dotted,
    line width = 1pt
    ]
    table {Data/ErgodicRate/CapacityRegion/CapacityFF/x5.dat};
    \addplot[
    black,
    mark = o,
    mark repeat = 2,
    mark size = 3,
    mark phase = 0,
    no marks,
    style = dotted,
    line width = 1pt
    ]
    table {Data/ErgodicRate/CapacityRegion/CapacityFF/y5.dat};
    \addplot[
    black,
    mark = o,
    mark repeat = 2,
    mark size = 3,
    mark phase = 0,
    no marks,
    style = dashed,
    line width = 1pt
    ]
    table {Data/ErgodicRate/CapacityRegion/CapacityFF/x2.dat};
    \addplot[
    black,
    mark = o,
    mark repeat = 2,
    mark size = 3,
    mark phase = 0,
    no marks,
    style = dashed,
    line width = 1pt
    ]
    table {Data/ErgodicRate/CapacityRegion/CapacityFF/y2.dat};
    \addplot[only marks, mark=*] coordinates {(5.9554152,5.8919879) (5.1781812, 5.3860099)};
    \node[] at (axis cs: 4.5,5){$(C_a^*,C_b^*)$};
    \node[] at (axis cs: 6,5.4){($A',B'$)};
    \node[] at (axis cs: 3,6.4){$\mathrm{EFF}_2^{\mathrm{DU-RSMA}} = \frac{\text{Area A}}{\text{Area B}}=\frac{C_a^*C_b^*}{A'B'}$};
\path[name path=lower1] (axis cs:0, 0) -- (axis cs: 5.1781812, 0);
\path[name path=upper1] (axis cs:0, 5.3860099) -- (axis cs: 5.1781812, 5.3860099);
\addplot[pattern=north west lines,pattern color=black]
fill between[
of = lower1 and upper1,
soft clip = {domain=0:5.1781812}
];
\path[name path=lower2] (axis cs:0, 0) -- (axis cs: 5.9554152, 0);
\path[name path=upper2] (axis cs:0, 5.8919879) -- (axis cs: 5.9554152, 5.8919879);
\addplot[pattern=dots,pattern color=black]
fill between[
of = lower2 and upper2,
soft clip = {domain=0:5.9554152}
];

.
\end{axis}
   \end{tikzpicture}
   \caption{Achievable ergodic capacity region.}
   \vspace{-2mm}
   \label{Achievable ER}
\end{figure}

In Fig. \ref{Achievable ER}, the ER of user $b$ is plotted against the ER of user $a$ for different values of the fraction $\alpha$ of $P_a$ allocated to the transmission of message $x_{1a}$, while the distances between the users and each RRH are the same as in Fig. \ref{Capacity Region}. This figure shows the achievable ergodic capacity region for three different cases. Specifically, the cases when only RRH $i$ is available for transmission, when only RRH $j$ is available for transmission, and finally when users can transmit to both RRHs and the optimal one to decode each message is selected using the selection criterion proposed in this work. The significant gain in the achievable ergodic capacity region should be highlighted. Furthermore, unlike the standard capacity regions where the corner points are connected by a straight line, as presented in the cases for RRHs $i$ and $j$, in the proposed scheme they are connected by a curved line, which allows us to obtain a wider capacity region. 
Taking this into account, we can also define the ergodic fill factor for the different values of $q$. Indicatively, the expression for $q=2$ is provided below 
\begin{equation}
\label{ergodic fill factor}
\begin{aligned}
&\mathrm{EFF}_2=\\
&\frac{\Bar{C}_a^*\Bar{C}_b^*}{\underbrace{\mathbb{E}\{\log_2(1+\max(\gamma_{a,i},\gamma_{a,j}))\}}_{A'}\underbrace{\mathbb{E}\{\log_2(1+\max(\gamma_{b,i},\gamma_{b,j}))\}}_{B'}},
\end{aligned}
\end{equation}
where $\Bar{C}_a^*$ and $\Bar{C}_b^*$ are the pair of ERs that maximizes the area of the largest rectangle that fits within the achievable ergodic capacity region. Specifically, for these simulation parameters, $\mathrm{EFF} = 0.795$.

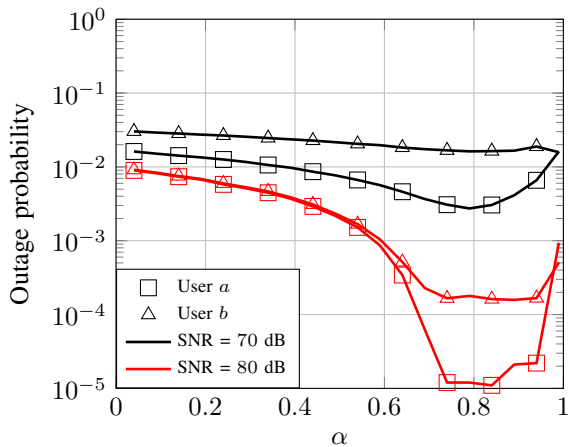
\begin{figure}
\centering
\begin{tikzpicture}
   \begin{semilogyaxis}[
   width = 0.85\linewidth,
   xlabel = {$\alpha$},
   ylabel = {Outage probability},
   ymin = 0.00001,
   ymax = 1,
   xmin = 0,
   xmax = 1,
   grid = major,
   legend entries = {{User $a$},{User $b$},{},{},{SNR = $70$ dB},{},{SNR = $80$ dB}},
   legend cell align = {left},
   legend style = {font = \scriptsize},
   legend style={at={(0,0)},anchor=south west}
   ]
   \addplot[
    black,
    mark = square,
    mark repeat = 2,
    mark size = 3,
    mark phase = 0,
    only marks,
    ]
    table {Data/Outage/OutagevsA/NOMARSMAA70.dat};
    \addplot[
    black,
    mark = triangle,
    mark repeat = 2,
    mark size = 3,
    mark phase = 0,
    only marks,
    ]
    table {Data/Outage/OutagevsA/NOMARSMAB70.dat};
    \addplot[
    red,
    mark = square,
    mark repeat = 2,
    mark size = 3,
    mark phase = 0,
    only marks,
    ]
    table{Data/Outage/OutagevsA/NOMARSMAA80.dat};
    \addplot[
    red,
    mark = triangle,
    mark repeat = 2,
    mark size = 3,
    mark phase = 0,
    only marks,
    ]
    table{Data/Outage/OutagevsA/NOMARSMAB80.dat};
    \addplot[
    black,
    mark = square,
    mark repeat = 2,
    mark size = 3,
    mark phase = 0,
    no marks,
    line width = 1pt
    ]
    table {Data/Outage/OutagevsA/NOMARSMAA70.dat};
    \addplot[
    black,
    mark = triangle,
    mark repeat = 2,
    mark size = 3,
    mark phase = 0,
    no marks,
    line width = 1pt
    ]
    table {Data/Outage/OutagevsA/NOMARSMAB70.dat};
    \addplot[
    red,
    mark = square,
    mark repeat = 2,
    mark size = 3,
    mark phase = 0,
    no marks,
    line width = 1pt
    ]
    table{Data/Outage/OutagevsA/NOMARSMAA80.dat};
    \addplot[
    red,
    mark = triangle,
    mark repeat = 2,
    mark size = 3,
    mark phase = 0,
    no marks,
    line width = 1pt
    ]
    table{Data/Outage/OutagevsA/NOMARSMAB80.dat};
   \end{semilogyaxis}
\end{tikzpicture}
\caption{Outage vs $\alpha$.}
\vspace{-4mm}
\label{OutagevsA}
\end{figure}

In Fig. \ref{OutagevsA}, the OP of each user is plotted against the fraction of power $P_a$ used to transmit the message $x_{1a}$, $\alpha$, for different SNRs. The lines shown in this figure are for $d_{ia}=5$ m, $d_{ja}=15$ m, $d_{ib} = 20$ m, $d_{ib} = 10$ m, and $R_a = R_b = 1.5$. As expected, increasing the SNR results in lower OP because the conditions under which the system operates are improved. It should be noted that user $a$ outperforms user $b$. This can be explained by the fact that user $a$ is closer to the RRHs than user $b$. $d_{ia}$ should be compared to $d_{jb}$ and $d_{ib}$ to $d_{ja}$. Note that the optimal $\alpha$ decreases as the SNR decreases.  
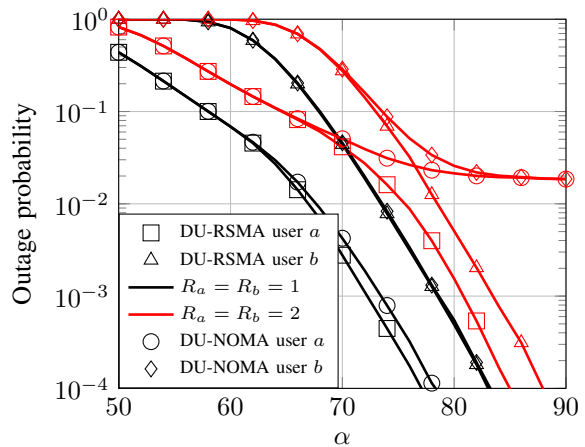
\begin{figure}
\centering
\begin{tikzpicture}
   \begin{semilogyaxis}[
   width = 0.85\linewidth,
   xlabel = {$\alpha$},
   ylabel = {Outage probability},
   ymin = 0.0001,
   ymax = 1,
   xmin = 50,
   xmax = 90,
   grid = major,
   legend entries = {{DU-RSMA user $a$},{DU-RSMA user $b$},{},{},{$R_a = R_{b} = 1$},{},{$R_a = R_{b}=2$},{},{DU-NOMA user $a$},{DU-NOMA user $b$}},
   legend cell align = {left},
   legend style = {font = \scriptsize},
   legend style={at={(0,0)},anchor=south west}
   ]
   \addplot[
    black,
    mark = square,
    mark repeat = 2,
    mark size = 3,
    mark phase = 0,
    only marks,
    ]
    table {Data/Outage/OutagevsSNR/NOMARSMAFinalARate1.dat};
    \addplot[
    black,
    mark = triangle,
    mark repeat = 2,
    mark size = 3,
    mark phase = 0,
    only marks,
    ]
    table {Data/Outage/OutagevsSNR/NOMARSMAFinalBRate1.dat};
    \addplot[
    red,
    mark = square,
    mark repeat = 2,
    mark size = 3,
    mark phase = 0,
    only marks,
    ]
    table{Data/Outage/OutagevsSNR/NOMARSMAFinalARate2.dat};
    \addplot[
    red,
    mark = triangle,
    mark repeat = 2,
    mark size = 3,
    mark phase = 0,
    only marks,
    ]
    table{Data/Outage/OutagevsSNR/NOMARSMAFinalBRate2.dat};
    \addplot[
    black,
    mark = square,
    mark repeat = 2,
    mark size = 3,
    mark phase = 0,
    no marks,
    line width = 1pt
    ]
    table {Data/Outage/OutagevsSNR/NOMARSMAFinalARate1.dat};
    \addplot[
    black,
    mark = triangle,
    mark repeat = 2,
    mark size = 3,
    mark phase = 0,
    no marks,
    line width = 1pt
    ]
    table {Data/Outage/OutagevsSNR/NOMARSMAFinalBRate1.dat};
    \addplot[
    red,
    mark = square,
    mark repeat = 2,
    mark size = 3,
    mark phase = 0,
    no marks,
    line width = 1pt
    ]
    table{Data/Outage/OutagevsSNR/NOMARSMAFinalARate2.dat};
    \addplot[
    red,
    mark = triangle,
    mark repeat = 2,
    mark size = 3,
    mark phase = 0,
    no marks,
    line width = 1pt
    ]
    table{Data/Outage/OutagevsSNR/NOMARSMAFinalBRate2.dat};
    \addplot[
    black,
    mark = o,
    mark repeat = 2,
    mark size = 3,
    mark phase = 0,
    only marks,
    ]
    table {Data/Outage/OutagevsSNR/NOMAFinalARate1.dat};
    \addplot[
    black,
    mark = diamond,
    mark repeat = 2,
    mark size = 3,
    mark phase = 0,
    only marks,
    ]
    table {Data/Outage/OutagevsSNR/NOMAFinalBRate1.dat};
    \addplot[
    red,
    mark = o,
    mark repeat = 2,
    mark size = 3,
    mark phase = 0,
    only marks,
    ]
    table{Data/Outage/OutagevsSNR/NOMAFinalARate2.dat};
    \addplot[
    red,
    mark = diamond,
    mark repeat = 2,
    mark size = 3,
    mark phase = 0,
    only marks,
    ]
    table{Data/Outage/OutagevsSNR/NOMAFinalBRate2.dat};
    \addplot[
    black,
    mark = square,
    mark repeat = 2,
    mark size = 3,
    mark phase = 0,
    no marks,
    line width = 1pt
    ]
    table {Data/Outage/OutagevsSNR/NOMAFinalARate1.dat};
    \addplot[
    black,
    mark = triangle,
    mark repeat = 2,
    mark size = 3,
    mark phase = 0,
    no marks,
    line width = 1pt
    ]
    table {Data/Outage/OutagevsSNR/NOMAFinalBRate1.dat};
    \addplot[
    red,
    mark = square,
    mark repeat = 2,
    mark size = 3,
    mark phase = 0,
    no marks,
    line width = 1pt
    ]
    table{Data/Outage/OutagevsSNR/NOMAFinalARate2.dat};
    \addplot[
    red,
    mark = triangle,
    mark repeat = 2,
    mark size = 3,
    mark phase = 0,
    no marks,
    line width = 1pt
    ]
    table{Data/Outage/OutagevsSNR/NOMAFinalBRate2.dat};
   \end{semilogyaxis}
\end{tikzpicture}
\caption{Outage vs transmit SNR.}
\vspace{-4mm}
\label{OutagevsSNR}
\end{figure}

In Fig. \ref{OutagevsSNR}, the performance of the proposed scheme is compared to that of DU-NOMA in terms of OP for various rates and for $d_{ia} = 5$ m, $d_{ib} = 25$ m, $d_{ja} = 30$ m, and $d_{jb} = 20$ m. As expected, increasing the SNR leads to reduced OP in every scenario. It is noted, that the OP of user $a$ is lower than the OP of user $b$, in both schemes. This is due to the fact that user $a$ is closer to the RRHs than user $b$, and thus it has less attenuation because of the path losses. The most important conclusion that can be extracted from this figure, is that the proposed scheme, i.e., DU-RSMA does not have an OP floor regardless of the values of the thresholds for correct message transmission. This is a significant advantage over DU-NOMA, which is known to have an OP floor for $\theta_1\theta_2 > 1$.
   
\section{Conclusions}
In this paper, a DU-RSMA scheme was proposed and investigated. Specifically, the performance of a system with two users and two RRHs was studied, assuming that the RRHs can decode each transmitted message independently and exchange information through an error-free feedback link. The performance of the system was investigated under the assumption of adaptive and fixed transmission rates. For the case of adaptive rates, to improve the system performance, we also proposed a selection criterion based on which each message is decoded by the optimal RRH. A new metric to evaluate the performance of MAC schemes, termed as fill factor, was also introduced. Closed-form expressions for the ER and OP of each user were derived and validated by simulation results. The simulations also showed the impact of each parameter on the system performance, and proved that the proposed scheme outperforms both NOMA and DU-RSMA without feedback link in terms of ER and OP. Finally, it is shown that the achievable capacity region is extended when this scheme is implemented.

\appendices
\section{Proof of Theorem \ref{th3}}
\label{Proof3}
In the proposed scheme, both users transmit their messages by sharing the same resource block.  Taking this into account, the ER of user $a$ will be given by \begin{equation}
    \label{Ergodic Split Max}
    \begin{aligned}
    \Bar{C}_a &= \mathbb{E}\{\max\left(\log_2(1+\gamma_{1a,i}),\log_2(1+\gamma_{1a,j})\right)\} \\ &\qquad+ \mathbb{E}\{\max\left[\log_2(1+\gamma_{2a,i}),\log_2(1+\gamma_{2a,j})\right]\}.
    \end{aligned}
\end{equation}
To decide whether the transmitted message is decoded by RRH $i$ or RRH $j$, the criterion presented in Section \ref{Proposed} is considered. To derive closed-form expressions for $P_1$ and $P_2$, a high-SNR approximation is assumed. Specifically, since in high-SNR regions the SINR of each message is given by
\begin{equation}
\label{SNR1a,i HIGH}
    \gamma_{1a,k}^H = \frac{\alpha l_{ka}\gamma_a\lvert h_{ka}\rvert^2}{(1-\alpha)l_{ka}\gamma_a\lvert h_{ka}\rvert^2+l_{kb}\gamma_b\lvert h_{kb}\rvert^2},
\end{equation}
\begin{equation}
\label{SNRb,i HIGH}
    \gamma_{b,k}^H = \frac{l_{kb}\gamma_b\lvert h_{kb}\rvert^2}{(1-\alpha)l_{ka}\gamma_a\lvert h_{ka}\rvert^2}
\end{equation}
and
\begin{equation}
\label{SNR2a,i HIGH}
    \gamma_{2a,k}^H = (1-\alpha)l_{ka}\gamma_a\lvert h_{ka}\rvert^2,
\end{equation}
for RRH $i$ and $j$, depending on the value of $k$. Message $x_{1a}$ is decoded by RRH $i$ when $\gamma_{1a,i}^H > \gamma_{1a,j}^H$, and message $x_{2a}$ is decoded by RRH $i$ when $\gamma_{2a,i}^H > \gamma_{2a,j}^H$. Otherwise, RRH $j$ decodes the transmitted message.
To derive $P_1$, starting from $\Pr\left(\gamma_{1a,i}^H > \gamma_{1a,j}^H\right)$, after some algebraic manipulations we get
\begin{equation}
\label{Channel Comparison 1}
\begin{aligned}
    \Pr\left(\gamma_{1a,i}^H > \gamma_{1a,j}^H\right)=\Pr\left(\lvert h_{ia}\rvert^2\lvert h_{jb}\rvert^2 > \frac{l_{ja}l_{ib}}{l_{ia}l_{jb}}\lvert h_{ja}\rvert^2\lvert h_{ib}\rvert^2\right).
    \end{aligned}
\end{equation}
To calculate the probability in \eqref{Channel Comparison 1}, considering that $\lvert h_{kl}\rvert^2$ follows an exponential distribution, we need to use the probability density function (PDF) of the product of two independent exponentially distributed random variables, which is given as
\begin{equation}
\label{PDF Product}
    f_Z(z) = 2\lambda_{x}\lambda_{y}\mathrm{K_0}\left(2\sqrt{\lambda_x\lambda_yz}\right), 
\end{equation}
where $Z = XY$, with $X,Y$ being exponentially distributed random variables with scale parameters $\lambda_x$ and $\lambda_y$, respectively, and $\mathrm{K_0}(\cdot)$ denotes the modified Bessel function of the second kind of order 0. 
Taking \eqref{PDF Product} and \eqref{Channel Comparison 1} into consideration, $P_1$ can be calculated as
\begin{equation}
\label{Pr11}
\begin{aligned}
    P_1 &= \int_0^\infty\int_{\frac{l_{ja}l_{ib}y}{l_{ia}l_{jb}}}^\infty4 \mathrm{K_0}\left(2\sqrt{ y}\right)\mathrm{K_0}\left(2\sqrt{x}\right)\mathrm{d}x\mathrm{d}y.
    \end{aligned}
\end{equation}
By implementing \cite[(1.12.1.5)]{prudnikov}, \eqref{Pr11} can be transformed into
\begin{equation}
    \begin{aligned}
        P_1 &= \int_0^\infty 4 \mathrm{K_0}\left(2\sqrt{ y}\right)\sqrt{\frac{l_{ja}l_{ib}y}{l_{jb}l_{ia}}}\mathrm{K_1}\left(2\sqrt{\frac{y}{l_{ia}l_{jb}}}\right)\mathrm{d}y.
    \end{aligned}
\end{equation}
Finally, by setting $x=\sqrt{y}$ and invoking \cite[2.16.33.1)]{prudnikov} for $A_{1,0}^3$, the final expression for $P_1$ is derived.

For $P_2$, starting from $\gamma_{2a,i}^H > \gamma_{2a,j}^H$ and after some algebraic manipulations, we derive
\begin{equation}
    P_2 = \Pr\left(\lvert h_{ia} \rvert^2 > \frac{l_{ja}}{l_{ia}}\lvert h_{ja} \rvert^2\right),
\end{equation}
which, as already stated, are random variables following the exponential distribution. Leveraging the PDF of a random variable (RV) following the exponential distribution, this probability can be easily calculated as
\begin{equation}
    P_2 = \int_0^\infty\int_\frac{l_{ja}y}{l_{ia}}^\infty e^{-x}e^{-y}\mathrm{d}x\mathrm{d}y.
\end{equation}
Calculating the integral gives the final expression for $P_2$.
As for $\Bar{C}_{1a}^i$, it is known that
\begin{equation}
\label{Capacity 1a,i}
\begin{aligned}
    \Bar{C}_{1a}^i &= \int_0^\infty \log_2(1+\gamma)f_{\gamma_{1a,i}}(\gamma)\mathrm{d}\gamma\\
    &=\frac{1}{\ln2}\int_0^\frac{\alpha}{1-\alpha}\frac{1-F_{\gamma_{1a,i}}(x)}{1+x}\mathrm{d}x.
    \end{aligned}
\end{equation}
To continue, it is obvious that the calculation of $F_{\gamma_{1a,i}}(x)$ is necessary, which is given by
\begin{equation}
\label{CDF 1a,i}
\begin{aligned}
    &F_{\gamma_{1a,i}}(x) = \Pr(\gamma_{1a,i} < x) \\
    &= \Pr\left(\frac{\alpha l_{ia}\gamma_a\lvert h_{ia}\rvert^2}{(1-\alpha)l_{ia}\gamma_a\lvert h_{ia}\rvert^2+l_{ib}\gamma_b\lvert h_{ib}\rvert^2+1} < x\right)  \\
  &=\begin{cases}
    1-\frac{ l_{ia}\gamma_a(\alpha-x(1-\alpha))e^{-\frac{x}{ l_{ia}\gamma_a(\alpha-x(1-\alpha))}}}{ l_{ia}\gamma_a(\alpha-x(1-\alpha))+l_{ib}\gamma_bx}, & \frac{a}{1-a} > x \\
    1, & \text{otherwise}.
    \end{cases}
\end{aligned}
\end{equation}
To calculate the probability when $\frac{a}{1-a} > x$, it follows that
\begin{equation}
\label{Proof_ia}
\begin{aligned}
    F_{\gamma_{1a,i}}\!(x)\! &= \int_0^{\infty}F_{\lvert h_{ia} \rvert^2}\left(\frac{xl_{ib}\gamma_by+x}{l_{ia}\gamma_a(a-x(1-a))}\right)f_{\lvert h_{ib} \rvert^2}(y)dy \\
    &= \!1\! -\! e^{-\frac{x}{l_{ia}\gamma_a(a-x(1-a))}}\!\!\int_0^{\infty}\!e^{-y\left(\frac{xl_{ib}\gamma_b}{ l_{ia}\gamma_a(a-x(1-a))}+1\right)}dy. 
\end{aligned}
\end{equation}
In \eqref{Proof_ia}, $F_Z(\cdot)$ and $f_Z(\cdot)$ denote the cumulative distribution function and PDF, respectively, of a random variable $Z$.
Using \eqref{CDF 1a,i} in \eqref{Capacity 1a,i}, \eqref{Capacity 1a,i} becomes
\begin{equation}
\label{Capacity 1a,i2}
    \begin{aligned}
    \Bar{C}_{1a}^i = \frac{1}{\ln2}\int_0^\frac{\alpha}{1-\alpha} \!\! \frac{e^{-\frac{x}{ l_{ia}\gamma_a(\alpha-x(1-\alpha))}}}{(1+x)l_{ib}\gamma_b\!\left(\frac{1}{l_{ib}\gamma_b}+\frac{x}{ l_{ia}\gamma_a(\alpha-x(1-\alpha))}\right)}\mathrm{d}x.
    \end{aligned}
\end{equation}
By setting $y=\frac{x}{\alpha-x(1-\alpha)}$ in \eqref{Capacity 1a,i2}, we get
\begin{equation}
\label{Capacity 1a,i3}
    \begin{aligned}
    \Bar{C}_{1a}^i = \frac{\alpha l_{ia}\gamma_a}{\ln2}\!\int_0^\infty \! \! \frac{e^{-\frac{y}{ l_{ia}\gamma_a}}}{(1+y)(1+y(1-\alpha))( l_{ia}\gamma_a+l_{ib}\gamma_by)}\mathrm{d}y.
    \end{aligned}
\end{equation}
The closed-form expression for $\Bar{C}_{1a}^i$ is derived by applying partial fraction decomposition in \eqref{Capacity 1a,i3}, where the different cases presented in \eqref{Ergodic 1A} correspond to the different forms of the denominator.
Finally, to calculate $\Bar{C}_{2a}^i$, we start from
\begin{equation}
\label{Capacity 2a,i}
    \Bar{C}_{2a}^i = \int_0^\infty\frac{1-F_{\gamma_{2a,i}}(x)}{1+x}\mathrm{d}x.
\end{equation}
Consequently, we need to calculate $F_{\gamma_{2a,i}}$ as follows
\begin{equation}
\label{CDF x2a,i}
    \begin{aligned}
        F_{\gamma_{2a,i}} &= \Pr\left(\gamma_{2a,i} < x\right) = \Pr\left((1-\alpha)l_{ia}\gamma_a\lvert h_{ia} \rvert^2 < x\right) \\
        &= \Pr\left(\lvert h_{ia} \rvert^2 < \frac{x}{(1-\alpha)l_{ia}\gamma_a}\right)= 1-e^{-\frac{x}{(1-\alpha) l_{ia}\gamma_a}}.
    \end{aligned}
\end{equation}
Substituting \eqref{CDF x2a,i} in \eqref{Capacity 2a,i}, we get
\begin{equation}
    \label{Capacity 2a,i2}
    \Bar{C}_{2a}^i = \frac{1}{\ln2}\int_0^\infty\frac{e^{-\frac{x}{(1-\alpha) l_{ia}\gamma_a}}}{1+x}\mathrm{d}x.
\end{equation}
Invoking \cite[(3.352.4)]{gradshteyn2014table} in \eqref{Capacity 2a,i2}, the final expression for $\Bar{C}_{2a}^i$ is derived. Following a similar procedure, the ER for messages $(x_{1a},x_{2a})$ can be derived for RRH $j$ completing the proof.

\section{Proof of Theorem \ref{th1}}
\label{Proof1}
In the proposed scheme, an outage for user $a$ occurs either when its message cannot be successfully decoded when NOMA is implemented, or when one of the two parts of the message (i.e., $x_{1a}, x_{2a}$) is lost when RSMA is implemented. It is assumed that to proceed to messages (i.e., $x_{b}, x_{2a}$), all previous messages have been correctly decoded. Since the RRHs cooperate via a feedback link, both RRHs must be unable to decode the transmitted messages, for an outage to occur, which means that the received SINRs at each RRH given by \eqref{SNR1a,i}-\eqref{SNR-NOMA-ab-a} is lower than the respective threshold. 
Taking this into account, there are 16 cases, which cover every possible scenario of outage for user $a$. These cases are a combination of $G_1^k-G_9^k$, as presented in \eqref{P_a}, given by
\begin{equation}
    G_1^k = \Pr\left(\gamma_{a,k}^{ab} < r_a,\gamma_{b,k}^{ba} < r_b,\gamma_{1a,k} < r_{a1}\right)
\end{equation}
\begin{equation}
    G_2^k = \Pr\left(\gamma_{a,k}^{ab} < r_a,\gamma_{b,k}^{ba} < r_b,\gamma_{1a,k} > r_{a1},\gamma_{b,k} < r_b\right)
\end{equation}
\begin{equation}
    G_3^k = \Pr\left(\gamma_{a,k}^{ab} < r_a,\gamma_{b,k}^{ba} < r_b,\gamma_{1a,k} < r_{a1},\gamma_{b,k} < r_b\right)
\end{equation}
\begin{equation}
\begin{aligned}
    G_4^k = &\Pr\left(\gamma_{a,k}^{ab} < r_a,\gamma_{b,k}^{ba} < r_b,\gamma_{1a,k} > r_{a1}, \right. \\ & \left. \qquad \gamma_{b,k} > r_b,\gamma_{2a,k} < r_{a2}\right)
    \end{aligned}
\end{equation}
\begin{equation}
\begin{aligned}
    G_5^k = & \Pr\left(\gamma_{a,k}^{ab} < r_a,\gamma_{b,k}^{ba} < r_b,\gamma_{1a,k} > r_{a1}, \right. \\ & \left. \qquad \gamma_{b,k} < r_b, \gamma_{2a,k} < r_{a2}\right)
    \end{aligned}
\end{equation}
\begin{equation}
\begin{aligned}
    G_6^k = &\Pr\left(\gamma_{a,k}^{ab} < r_a,\gamma_{b,k}^{ba} < r_b,\gamma_{1a,k} < r_{a1}, \right. \\ & \left. \qquad \gamma_{b,k} > r_b,\gamma_{2a,k} < r_{a2}\right)
    \end{aligned}
\end{equation}
\begin{equation}
\begin{aligned}
    G_7^k = & \Pr\left(\gamma_{a,k}^{ab} < r_a,\gamma_{b,k}^{ba} < r_b,\gamma_{1a,k} < r_{a1}, \right. \\ & \left. \qquad \gamma_{b,k} < r_b,\gamma_{2a,k} < r_{a2}\right)
    \end{aligned}
\end{equation}
\begin{equation}
    G_8^k = \Pr\left(\gamma_{a,k}^{ba} < r_a,\gamma_{b,k}^{ba} > r_b\right)
\end{equation}
\begin{equation}
    G_9^k = \Pr\left(\gamma_{a,k}^{ba} < r_a,\gamma_{b,k}^{ba}<r_b\right).
\end{equation}
More importantly, the cases in \eqref{P_a} are disjoint.

Regarding  the fifth term of \eqref{P_a}, we need to calculate $G_4^iG_4^j$. Since the individual terms are symmetrical for each RRH, we only need to calculate one of them to derive the final expression. Indicatively, we present the procedure for calculating $G_4^i$. 
Invoking \eqref{SNR1a,i}-\eqref{SNR2a,i}, \eqref{SNR-NOMA-ab-a}, and \eqref{SNR-NOMA-ab-b} in the expression for $\Pr(A_{5i})$ and after some algebraic manipulations, the following inequalities arise for each individual event
\begin{equation}
\label{gamma a,i^abProof2}
    \lvert h_{ia} \rvert^2 < \frac{\theta_1l_{ib}\gamma_b\lvert h_{ib}\rvert^2+\theta_1}{l_{ia}\gamma_a},
\end{equation}
\begin{equation}
\label{gamma b,i^baProof2}
    \lvert h_{ia} \rvert^2 > \frac{l_{ib}\gamma_b\lvert h_{ib}\rvert^2-\theta_2}{\theta_2l_{ia}\gamma_a} ,
\end{equation}
\begin{equation}
\label{gamma 1a,i}
\begin{aligned}
    \lvert h_{ia} \rvert^2 > \frac{\theta_{11}l_{ib}\gamma_b\lvert h_{ib}\rvert^2+\theta_{11}}{l_{ia}\gamma_a(\alpha-\theta_{11}(1-\alpha))}, \quad \frac{\alpha}{1-\alpha} > \theta_{11},
    \end{aligned}
\end{equation}
\begin{equation}
\label{gamma b,i}
    \lvert h_{ia} \rvert^2 < \frac{l_{ib}\gamma_b\lvert h_{ib}\rvert^2-\theta_2}{\theta_2l_{ia}\gamma_a(1-\alpha)},
\end{equation}
\begin{equation}
\label{gamma 2a,i}
    \lvert h_{ia} \rvert^2 < \frac{\theta_{12}}{l_{ia}\gamma_a(1-\alpha)}.
\end{equation}
Since the right-hand side terms are not constant, it is necessary to check the relationships between them. Specifically, considering \eqref{gamma b,i^baProof2} and \eqref{gamma 1a,i}, we have
\begin{equation}
    \begin{cases}
        \frac{l_{ib}\gamma_b\lvert h_{ib}\rvert^2-\theta_2}{\theta_2l_{ia}\gamma_a} \! > \! \frac{\theta_{11}l_{ib}\gamma_b\lvert h_{ib}\rvert^2+\theta_{11}}{l_{ia}\gamma_a(\alpha-\theta_{11}(1-\alpha))},  \lvert h_{ib}\rvert^2 \! > \! f_6(l_{ib}) \cap F_2 \! < \! 0 \\
        \frac{l_{ib}\gamma_b\lvert h_{ib}\rvert^2-\theta_2}{\theta_2l_{ia}\gamma_a} \! < \! \frac{\theta_{11}l_{ib}\gamma_b\lvert h_{ib}\rvert^2+\theta_{11}}{l_{ia}\gamma_a(\alpha-\theta_{11}(1-\alpha))},  \text{otherwise.}
    \end{cases}
\end{equation}
Next, considering \eqref{gamma a,i^abProof2} and \eqref{gamma b,i}, we have
\begin{equation}
    \begin{cases}
        \frac{\theta_1l_{ib}\gamma_b\lvert h_{ib}\rvert^2+\theta_1}{l_{ia}\gamma_a} \! < \! \frac{l_{ib}\gamma_b\lvert h_{ib}\rvert^2-\theta_2}{\theta_2l_{ia}\gamma_a(1-\alpha)}, \lvert h_{ib}\rvert^2 \cap 1\!-\!\theta_1\theta_2(1-\alpha) \! > \! 0 \\
        \frac{\theta_1l_{ib}\gamma_b\lvert h_{ib}\rvert^2+\theta_1}{l_{ia}\gamma_a} \! > \! \frac{l_{ib}\gamma_b\lvert h_{ib}\rvert^2-\theta_2}{\theta_2l_{ia}\gamma_a(1-\alpha)}, \text{otherwise.}
    \end{cases}
\end{equation}
Finally, comparing the right-hand side of \eqref{gamma a,i^abProof2} and \eqref{gamma 2a,i}, we get
\begin{equation}
    \begin{cases}
        \frac{\theta_1l_{ib}\gamma_b\lvert h_{ib}\rvert^2+\theta_1}{l_{ia}\gamma_a} < \frac{\theta_{12}}{l_{ia}\gamma_a(1-\alpha)}, &\lvert h_{ib}\rvert^2 <f_5(l_{ib}) \\
        \frac{\theta_1l_{ib}\gamma_b\lvert h_{ib}\rvert^2+\theta_1}{l_{ia}\gamma_a} > \frac{\theta_{12}}{l_{ia}\gamma_a(1-\alpha)}, &\lvert h_{ib}\rvert^2 >f_5(l_{ib}).
    \end{cases}
\end{equation}
Assuming that $F_2<0$, $1-\theta_1\theta_2(1-\alpha) > 0$, and $\max\left(f_4(l_{ib}),f_6(l_{ib})\right)<f_5(l_{ib})$, $\Pr(A_{5i})$ is given as
\begin{equation}
\label{A_5i}
\begin{aligned}
    &G_{4}^i = \\&\Pr\left(\frac{l_{ib}\gamma_b\lvert h_{ib}\rvert^2-\theta_2}{\theta_2l_{ia}\gamma_a} < \lvert h_{ia}\rvert^2 < \frac{\theta_1l_{ib}\gamma_b\lvert h_{ib}\rvert^2+\theta_1}{l_{ia}\gamma_a}\right.,\\ &\left.\max\left(f_4(l_{ib}),f_6(l_{ib})\right)<\lvert h_{ib}\rvert^2<f_5(l_{ib})\right).
\end{aligned}  
\end{equation}
For \eqref{A_5i} to have non-zero values, it must hold
\begin{equation}
\begin{aligned}
     &\frac{l_{ib}\gamma_b\lvert h_{ib}\rvert^2-\theta_2}{\theta_2l_{ia}\gamma_a} <  \frac{\theta_1l_{ib}\gamma_b\lvert h_{ib}\rvert^2+\theta_1}{l_{ia}\gamma_a}.
\end{aligned}
\end{equation}
After some algebraic manipulations, we get
\begin{equation}
    \begin{cases}
        \frac{l_{ib}\gamma_b\lvert h_{ib}\rvert^2-\theta_2}{\theta_2l_{ia}\gamma_a} \! < \!  \frac{\theta_1l_{ib}\gamma_b\lvert h_{ib}\rvert^2+\theta_1}{l_{ia}\gamma_a}, 1-\theta_1\theta_2 < 0 \\
        \frac{l_{ib}\gamma_b\lvert h_{ib}\rvert^2-\theta_2}{\theta_2l_{ia}\gamma_a} \! < \!  \frac{\theta_1l_{ib}\gamma_b\lvert h_{ib}\rvert^2+\theta_1}{l_{ia}\gamma_a}, \lvert h_{ib}\rvert^2 \! < \! f_9(l_{ib}) \! \cap \! 1\!-\!\theta_1\theta_2 \!>\! 0
    \end{cases}
\end{equation}
Leveraging the CDF of $\lvert h_{ia}\rvert^2$ and the PDF of $\lvert h_{ib}\rvert^2$, it occurs
\begin{equation}
\begin{aligned}
\label{A_5i cases}
&G_{4}^i \\&=\!
    \begin{cases}
\int_{\max\left(f_4(l_{ib}),f_6(l_{ib})\right)}^{f_5(l_{ib})}F_{\lvert h_{ia}\rvert^2}\left(\frac{\theta_1l_{ib}\gamma_by+\theta_1}{l_{ia}\gamma_a}\right)f_{\lvert h_{ib}\rvert^2}(y)\mathrm{d}y\\-\int_{\max\left(f_4(l_{ib}),f_6(l_{ib})\right)}^{f_5(l_{ib})}\!F_{\lvert h_{ia}\rvert^2}\!\left(\frac{l_{ib}\gamma_b\lvert h_{ib}\rvert^2-\theta_2}{\theta_2l_{ia}\gamma_a}\right)\!f_{\lvert h_{ib}\rvert^2}(y)\mathrm{d}y,
\\ \qquad \qquad \qquad \qquad \qquad \qquad \qquad \qquad \quad \ 1-\theta_1\theta_2 < 0 \\
\int_{\max\left(f_4(l_{ib}),f_6(l_{ib})\right)}^{\min\left(f_5(l_{ib}),f_9(l_{ib})\right)}F_{\lvert h_{ia}\rvert^2}\left(\frac{\theta_1l_{ib}\gamma_by+\theta_1}{l_{ia}\gamma_a}\right)f_{\lvert h_{ib}\rvert^2}(y)\mathrm{d}y\\-\int_{\max\left(f_4(l_{ib}),f_6(l_{ib})\right)}^{\min\left(f_5(l_{ib}),f_9(l_{ib})\right)}F_{\lvert h_{ia}\rvert^2}\!\left(\frac{l_{ib}\gamma_b\lvert h_{ib}\rvert^2-\theta_2}{\theta_2l_{ia}\gamma_a}\right)\!f_{\lvert h_{ib}\rvert^2}(y)\mathrm{d}y
,\\ \qquad \qquad \qquad \qquad \qquad \qquad \qquad \qquad \quad \ 1-\theta_1\theta_2 > 0.
    \end{cases}
    \end{aligned}
\end{equation}
Calculating \eqref{A_5i cases}, $G_{45}^i$ in Table \ref{case G4} is derived. Taking into account every possible case between the right-hand side terms of \eqref{gamma a,i^abProof2}-\eqref{gamma 2a,i} the rest of the terms of Table \ref{case G4} are derived. Following a similar procedure, the rest of the tables are calculated, which completes the proof.

\section{Proof of Theorem \ref{th2}}
\label{Proof2}
In order for an outage for user $b$ to occur, both RRHs must be unable to decode its message both when NOMA and RSMA are implemented. Considering this, the events that an outage occurs are 
a combination of $G_1^k-G_3^k$, 
\begin{equation}
    G_{10}^k = \Pr\left(\gamma_{a,k}^{ab} > r_a, \gamma_{b,k}^{ba} < r_b\right),
\end{equation}
and
\begin{equation}
    G_{11}^k = \Pr\left(\gamma_{a,k}^{ab} < r_a, \gamma_{b,k}^{ba}\right).
\end{equation}

Following a procedure similar to the previous proof, by substituting \eqref{SNR1a,i}-\eqref{SNR-NOMA-ab-b} at each event and leveraging the PDF of the RVs $\lvert h_{ia}\rvert^2$, $\lvert h_{ib}\rvert^2$, $\lvert h_{ja}\rvert^2$, $\lvert h_{jb}\rvert^2$, \eqref{Outage b} is calculated. This procedure is shown indicatively for $G_3^i$. 
\begin{equation}
    G_3^i = \Pr\left(\gamma_{a,i}^{ab} < r_a,\gamma_{b,i}^{ba} < r_b,\gamma_{1a,i} < r_{a1},\gamma_{b,i} < r_b\right).
\end{equation}
Specifically, considering each individual event, we have
\begin{equation}
\label{gamma a,i^ab}
    \lvert h_{ia} \rvert^2 < \frac{\theta_1l_{ib}\gamma_b\lvert h_{ib}\rvert^2+\theta_1}{l_{ia}\gamma_a},    
\end{equation}
\begin{equation}
\label{gamma b,i^ba}
    \lvert h_{ia} \rvert^2 > \frac{l_{ib}\gamma_b\lvert h_{ib}\rvert^2-\theta_2}{\theta_2l_{ia}\gamma_a},    
\end{equation}
\begin{equation}
\label{gamma 1a,i lower}
\begin{aligned}
    \lvert h_{ia} \rvert^2 < \frac{\theta_{11}l_{ib}\gamma_b\lvert h_{ib}\rvert^2+\theta_{11}}{l_{ia}\gamma_a(\alpha-\theta_{11}(1-\alpha))}, \quad \frac{\alpha}{1-\alpha} > \theta_{11},  
    \end{aligned}
\end{equation}
\begin{equation}
\label{gamma b,i lower}
    \lvert h_{ia} \rvert^2 > \frac{l_{ib}\gamma_b\lvert h_{ib}\rvert^2-\theta_2}{\theta_2l_{ia}\gamma_a(1-\alpha)}.
\end{equation}
Since \eqref{gamma a,i^ab}-\eqref{gamma b,i lower} must all hold, $G_3^i$ can be calculated as
\begin{equation}
\label{G_3 first}
\begin{aligned}
    G_3^i=&\max\left(\frac{l_{ib}\gamma_b\lvert h_{ib}\rvert^2-\theta_2}{\theta_2l_{ia}\gamma_a(1-\alpha)},\frac{l_{ib}\gamma_b\lvert h_{ib}\rvert^2-\theta_2}{\theta_2l_{ia}\gamma_a}\right) < \lvert h_{ia} \rvert^2< \\& \min\left(\frac{\theta_1l_{ib}\gamma_b\lvert h_{ib}\rvert^2+\theta_1}{l_{ia}\gamma_a},\frac{\theta_{11}l_{ib}\gamma_b\lvert h_{ib}\rvert^2+\theta_{11}}{l_{ia}\gamma_a(\alpha-\theta_{11}(1-\alpha))}\right).
    \end{aligned}
\end{equation}
Note, however, that since $0<1-\alpha<1$, $\frac{l_{ib}\gamma_b\lvert h_{ib}\rvert^2-\theta_2}{\theta_2l_{ia}\gamma_a(1-\alpha)} > \frac{l_{ib}\gamma_b\lvert h_{ib}\rvert^2-\theta_2}{\theta_2l_{ia}\gamma_a}$, and $\frac{\theta_{11}l_{ib}\gamma_b\lvert h_{ib}\rvert^2+\theta_{11}}{l_{ia}\gamma_a(\alpha-\theta_{11}(1-\alpha))}$ should be considered only if $\frac{\alpha}{1-\alpha} >\theta_{11}$. Thus, \eqref{G_3 first} becomes
\begin{equation}
\label{G_3 second}
    G_3^i \!=\!
    \begin{cases}
        \frac{l_{ib}\gamma_b\lvert h_{ib}\rvert^2-\theta_2}{\theta_2l_{ia}\gamma_a(1-\alpha)} < \lvert h_{ia} \rvert^2<  \min\left(\frac{\theta_1l_{ib}\gamma_b\lvert h_{ib}\rvert^2+\theta_1}{l_{ia}\gamma_a}, \right.\\
        \qquad \left.\frac{\theta_{11}l_{ib}\gamma_b\lvert h_{ib}\rvert^2+\theta_{11}}{l_{ia}\gamma_a(\alpha-\theta_{11}(1-\alpha))}\right),  \qquad \qquad \qquad \ \, \frac{\alpha}{1-\alpha} > \theta_{11}\\
        \frac{l_{ib}\gamma_b\lvert h_{ib}\rvert^2-\theta_2}{\theta_2l_{ia}\gamma_a(1-\alpha)} < \lvert h_{ia} \rvert^2<\frac{\theta_1l_{ib}\gamma_b\lvert h_{ib}\rvert^2+\theta_1}{l_{ia}\gamma_a},  \, \frac{\alpha}{1-\alpha} < \theta_{11}.
    \end{cases}
\end{equation}
In the first case, when $\frac{\theta_1l_{ib}\gamma_b\lvert h_{ib}\rvert^2+\theta_1}{l_{ia}\gamma_a} \lessgtr \frac{\theta_{11}l_{ib}\gamma_b\lvert h_{ib}\rvert^2+\theta_{11}}{l_{ia}\gamma_a(\alpha-\theta_{11}(1-\alpha))}$ should be calculated, while in the second case it can be calculated by using the PDF of $\lvert h_{ia}\rvert^2$, $\lvert h_{ib}\rvert^2$, on the condition that $\frac{l_{ib}\gamma_b\lvert h_{ib}\rvert^2-\theta_2}{\theta_2l_{ia}\gamma_a(1-\alpha)}<\frac{\theta_1l_{ib}\gamma_b\lvert h_{ib}\rvert^2+\theta_1}{l_{ia}\gamma_a}$. Extending the first case, \eqref{G_3 second} becomes
\begin{equation}
\label{G_3 third}
    G_{3}^i=
    \begin{cases}
        \frac{l_{ib}\gamma_b\lvert h_{ib}\rvert^2-\theta_2}{\theta_2l_{ia}\gamma_a(1-\alpha)} < \lvert h_{ia} \rvert^2<  \frac{\theta_{11}l_{ib}\gamma_b\lvert h_{ib}\rvert^2+\theta_{11}}{l_{ia}\gamma_a(\alpha-\theta_{11}(1-\alpha))},\\
        \qquad \qquad \qquad \qquad \ \theta_{11}-\theta_1(\alpha-\theta_{11}(1-\alpha))<0\\
        \frac{l_{ib}\gamma_b\lvert h_{ib}\rvert^2-\theta_2}{\theta_2l_{ia}\gamma_a(1-\alpha)} < \lvert h_{ia} \rvert^2<\frac{\theta_1l_{ib}\gamma_b\lvert h_{ib}\rvert^2+\theta_1}{l_{ia}\gamma_a},\\ \qquad \qquad \qquad \qquad \ \theta_{11}-\theta_1(\alpha-\theta_{11}(1-\alpha))>0.
    \end{cases}
\end{equation}
In both cases, we should check that the left-hand side of the inequality is less than the right-hand side of the inequality, and finally substitute the PDF of $\lvert h_{ia}\rvert^2$, $\lvert h_{ib}\rvert^2$ to calculate the probability. The closed-form expressions derived by the above procedure, along with their conditions, are presented in Table \ref{case G3}. Similarly, the probabilities of the remaining events can be calculated, completing the proof.

\bibliographystyle{ieeetr}
\bibliography{Bibliography}

\begin{thebibliography}{10}

\bibitem{proceedings}
N.~G. Evgenidis, N.~A. Mitsiou, V.~I. Koutsioumpa, S.~A. Tegos, P.~D. Diamantoulakis, and G.~K. Karagiannidis, ``Multiple access in the era of distributed computing and edge intelligence,'' {\em Proc. IEEE}, pp.~1--30, 2024.

\bibitem{6G}
Z.~Zhang, Y.~Xiao, Z.~Ma, M.~Xiao, Z.~Ding, X.~Lei, G.~K. Karagiannidis, and P.~Fan, ``6{G} wireless networks: Vision, requirements, architecture, and key technologies,'' {\em IEEE Veh. Technol. Mag.}, vol.~14, no.~3, pp.~28--41, 2019.

\bibitem{NGMA}
P.~D. Diamantoulakis, N.~D. Chatzidiamantis, A.~L. Moustakas, and G.~K. Karagiannidis, ``Next generation multiple access: Performance gains from uplink mimo-noma,'' {\em IEEE Open J. Commun. Soc.}, vol.~3, pp.~2298--2313, 2022.

\bibitem{NOMA}
B.~Makki, K.~Chitti, A.~Behravan, and M.-S. Alouini, ``A survey of {NOMA}: Current status and open research challenges,'' {\em IEEE Open J. Commun. Soc.}, vol.~1, pp.~179--189, 2020.

\bibitem{SANOMA}
S.~A. Tegos, P.~D. Diamantoulakis, A.~S. Lioumpas, P.~G. Sarigiannidis, and G.~K. Karagiannidis, ``Slotted {ALOHA} with {NOMA} for the next generation {IoT},'' {\em IEEE Trans. Commun.}, vol.~68, no.~10, pp.~6289--6301, 2020.

\bibitem{ChoiNOMA}
J.~Choi, ``{NOMA}-based random access with multichannel {ALOHA},'' {\em IEEE J. Sel Areas Commun.}, vol.~35, no.~12, pp.~2736--2743, 2017.

\bibitem{ApoNOMA}
A.~A. Tegos, S.~A. Tegos, D.~Tyrovolas, P.~D. Diamantoulakis, P.~Sarigiannidis, and G.~K. Karagiannidis, ``Breaking orthogonality in uplink with randomly deployed sources,'' {\em IEEE Open J. Commun. Soc.}, vol.~5, pp.~566--582, 2024.

\bibitem{RSMA}
Y.~Mao, O.~Dizdar, B.~Clerckx, R.~Schober, P.~Popovski, and H.~V. Poor, ``Rate-splitting multiple access: Fundamentals, survey, and future research trends,'' {\em IEEE Commun. Surveys Tuts.}, vol.~24, no.~4, pp.~2073--2126, 2022.

\bibitem{TSNOMA2}
D.~Tse and P.~Viswanath, {\em Fundamentals of wireless communication}.
\newblock Cambridge university press, 2005.

\bibitem{EncodingNOMA}
T.~Han and K.~Kobayashi, ``A new achievable rate region for the interference channel,'' {\em IEEE Trans. Inf. Theory}, vol.~27, no.~1, pp.~49--60, 1981.

\bibitem{RSMA_Begin}
B.~Rimoldi and R.~Urbanke, ``A rate-splitting approach to the {Gaussian} multiple-access channel,'' {\em IEEE Trans. Inf. Theory}, vol.~42, no.~2, pp.~364--375, 1996.

\bibitem{SumRateMaximization}
Z.~Yang, M.~Chen, W.~Saad, W.~Xu, and M.~Shikh-Bahaei, ``Sum-rate maximization of uplink rate splitting multiple access ({RSMA}) communication,'' {\em IEEE Trans. Mobile Comput.}, vol.~21, no.~7, pp.~2596--2609, 2022.

\bibitem{YueCognitive}
Y.~Xiao, S.~A. Tegos, P.~D. Diamantoulakis, Z.~Ma, and G.~K. Karagiannidis, ``On the ergodic rate of cognitive radio inspired uplink multiple access,'' {\em IEEE Commun. Lett.}, vol.~27, no.~1, pp.~95--99, 2023.

\bibitem{Tsiftsiscognitive}
H.~Liu, Z.~Bai, H.~Lei, G.~Pan, K.~J. Kim, and T.~A. Tsiftsis, ``A new rate splitting strategy for uplink {CR-NOMA} systems,'' {\em IEEE Trans. Veh. Technol.}, vol.~71, no.~7, pp.~7947--7951, 2022.

\bibitem{RSMAImperfectSIC}
A.~P. Chrysologou, S.~A. Tegos, P.~D. Diamantoulakis, N.~D. Chatzidiamantis, P.~C. Sofotasios, and G.~K. Karagiannidis, ``On the coexistence of heterogeneous services in 6{G} networks: An imperfection-aware {RSMA} framework,'' {\em IEEE Trans. Commun.}, pp.~1--1, 2024.

\bibitem{RSMA1order}
Y.~Zhu, X.~Wang, Z.~Zhang, X.~Chen, and Y.~Chen, ``A rate-splitting non-orthogonal multiple access scheme for uplink transmission,'' in {\em Proc. Int. Conf. Wireless Commun. Signal Process. ({WCSP})}, pp.~1--6, 2017.

\bibitem{DecodingOrder}
H.~Liu, T.~A. Tsiftsis, K.~J. Kim, K.~S. Kwak, and H.~V. Poor, ``Rate splitting for uplink {NOMA} with enhanced fairness and outage performance,'' {\em IEEE Trans. Wireless Commun.}, vol.~19, no.~7, pp.~4657--4670, 2020.

\bibitem{RSMAallorders}
S.~A. Tegos, P.~D. Diamantoulakis, and G.~K. Karagiannidis, ``On the performance of uplink rate-splitting multiple access,'' {\em IEEE Commun. Lett.}, vol.~26, no.~3, pp.~523--527, 2022.

\bibitem{RSMArandomusers}
H.~Lu, X.~Xie, Z.~Shi, H.~Lei, N.~Zhao, and J.~Cai, ``Outage performance of uplink rate splitting multiple access with randomly deployed users,'' {\em IEEE Trans. Wireless Commun.}, vol.~23, no.~2, pp.~1308--1326, 2024.

\bibitem{Pappi}
K.~N. Pappi, P.~D. Diamantoulakis, and G.~K. Karagiannidis, ``Distributed uplink-{NOMA} for cloud radio access networks,'' {\em IEEE Commun. Lett.}, vol.~21, no.~10, pp.~2274--2277, 2017.

\bibitem{DistributedNOMA}
P.~D. Diamantoulakis and G.~K. Karagiannidis, ``Performance analysis of distributed uplink {NOMA},'' {\em IEEE Commun. Lett.}, vol.~25, no.~3, pp.~788--792, 2021.

\bibitem{Alouini}
B.~Makki, K.~Chitti, A.~Behravan, and M.-S. Alouini, ``A survey of {NOMA}: Current status and open research challenges,'' {\em IEEE Open J. Commun. Soc.}, vol.~1, pp.~179--189, 2020.

\bibitem{Proportional_Fairness}
Z.-Q. Luo and S.~Zhang, ``Dynamic spectrum management: Complexity and duality,'' {\em IEEE J. Sel. Topics Signal Process.}, vol.~2, no.~1, pp.~57--73, 2008.

\bibitem{gradshteyn2014table}
I.~S. Gradshteyn and I.~M. Ryzhik, {\em Table of Integrals, Series, and Products}.
\newblock Academic Press, 2014.

\bibitem{prudnikov}
A.~P. Prudnikov, Y.~A. Brychkov, and O.~I. Marichev, {\em Integrals and Series: Special Functions}, vol.~2.
\newblock CRC Press, 1986.

\end{thebibliography}
\end{document}